\documentclass[twocolumn]{aastex63}

\usepackage{amsmath}
\usepackage[caption=false]{subfig}
\usepackage{graphicx}
\usepackage{xcolor}
\usepackage{txfonts}
\usepackage{rotating}

\def\etal{et al.\ }
\received{November 30, 2021}
\revised{March 4, 2022}
\accepted{March 23, 2022}

\submitjournal{ApJ}

\shorttitle{MIR studies of dusty sources in GC}
\shortauthors{Bhat et al.}
\graphicspath{{./}{figures/}}

\begin{document}

\title{Mid-Infrared studies of dusty sources in the Galactic Center}

\author[0000-0002-4408-0650]{Harshitha K. Bhat}
\affiliation{I.Physikalisches Institut der Universität zu Köln, Zülpicher Str. 77, D-50937 Köln, Germany}
\affiliation{Max-Planck-Institut für Radioastronomie (MPIfR), Auf dem Hügel 69, D-53121 Bonn, Germany}

\author[0000-0001-7134-9005]{Nadeen B. Sabha}
\affiliation{Institut für Astro- und Teilchenphysik, Universität Innsbruck, Technikerstrasse 25/8, A-6929 Innsbruck, Austria}
\affiliation{I.Physikalisches Institut der Universität zu Köln, Zülpicher Str. 77, D-50937 Köln, Germany}

\author[0000-0001-6450-1187]{Michal Zaja\v{c}ek}
\affiliation{Department of Theoretical Physics and Astrophysics, Faculty of Science, Masaryk University, Kotl\'a\v{r}sk\'a 2, 611 37 Brno, Czech Republic}
\affiliation{I.Physikalisches Institut der Universität zu Köln, Zülpicher Str. 77, D-50937 Köln, Germany}

\author[0000-0001-6049-3132]{Andreas Eckart}
\affiliation{I.Physikalisches Institut der Universität zu Köln, Zülpicher Str. 77, D-50937 Köln, Germany}
\affiliation{Max-Planck-Institut für Radioastronomie (MPIfR), Auf dem Hügel 69, D-53121 Bonn, Germany}

\author[0000-0001-5404-797X]{ Rainer Schödel}
\affiliation{Instituto de Astrofísica de Andalucía (CSIC), Glorieta de la Astronomía s/n, 18008 Granada, Spain}

\author[0000-0002-3004-6208]{S. Elaheh Hosseini}
\affiliation{I.Physikalisches Institut der Universität zu Köln, Zülpicher Str. 77, D-50937 Köln, Germany}
\affiliation{Max-Planck-Institut für Radioastronomie (MPIfR), Auf dem Hügel 69, D-53121 Bonn, Germany}

\author[0000-0002-9850-2708]{Florian Peißker}
\affiliation{I.Physikalisches Institut der Universität zu Köln, Zülpicher Str. 77, D-50937 Köln, Germany}

\author[0000-0001-7470-3321]{Anton Zensus}
\affiliation{Max-Planck-Institut für Radioastronomie (MPIfR), Auf dem Hügel 69, D-53121 Bonn, Germany}



\begin{abstract}

Mid-Infrared (MIR) images of the Galactic center show extended gas and dust features along with bright IRS sources.  Some of these dust features are a part of ionized clumpy streamers orbiting Sgr~A*, known as the mini-spiral. We present their proper motions over 12 year time period and report their flux densities in $N$-band filters {and derive their spectral indices}. The observations were carried out by VISIR at ESO VLT. High-pass filtering led to the detection of several resolved filaments and clumps along the mini-spiral. Each source was fit by a 2-D Gaussian profile to determine the offsets and aperture sizes. We perform aperture photometry to extract fluxes in two different bands. We present the proper motions of the largest consistent set of resolved and reliably determined sources. In addition to stellar orbital motions, we  identify a stream-like motion of extended clumps along the mini-spiral. We also detect MIR counterparts of the radio tail components of the IRS7 source. They show a clear kinematical deviation with respect to the star. They likely represent Kelvin-Helmholtz instabilities formed downstream in the shocked stellar wind. We also analyze the shape and the orientation of the extended late-type IRS3 star that is consistent with the ALMA sub-mm detection of the source.  Its puffed-up envelope with the radius of $\sim 2\times 10^6\,R_{\odot}$ could be the result of the red-giant collision with a nuclear jet, which was followed by the tidal prolongation along the orbit. 
\end{abstract}



\section{Introduction} \label{sec:intro}

 High-angular resolution observations of the vicinity of the compact radio source Sgr~A* in the Galactic center (GC), which is associated with the supermassive black hole \citep[SMBH;][]{Ecakrt2002MNRAS.331..917E,2010RvMP...82.3121G,2013CQGra..30x4003F,2017FoPh...47..553E}, showed the presence of the ring of dense clumpy molecular and neutral gas and warm dust called the circum-nuclear disc (CND), extending from $\sim 1.5$ to $\sim 7$ pc \citep{2004A&A...413..949V,Christopher2005ApJ...622..346C,2013ApJ...779...47M,Hsieh2021ApJ...913...94H}. The CND surrounds an ionized central cavity which has a much lower mean gas density within $\sim 1-1.5$ pc radius with the total mass of $\sim 60\,M_{\odot}$ \citep{Lo1983Natur.306..647L,2016MNRAS.459.1721B}. At the Bondi radius, $r_{\rm B}\sim 0.14\,(k_{\rm B}T/1.3\,{\rm keV})^{-1}{\rm pc}$, the mean electron number density is $n_{\rm e}=26\,f_{\rm V}^{-1/2}\,{\rm cm^{-3}}$ \citep{2003ApJ...591..891B}, where $f_{\rm V}$ is the filling factor of $1.3\,{\rm keV}$ plasma, while the CND density reaches $\sim 10^6-10^8\,{\rm cm^{-3}}$ in molecular cores \citep{Jackson1993ApJ...402..173J,2004ApJ...616..231S,Christopher2005ApJ...622..346C}. \citet{2018MNRAS.474.3787M} found a depression in the X-ray surface brightness at the position of the CND, which could be attributed to the CND acting as a barrier for the hot and diluted plasma in the central cavity. A system of orbiting ionized clumpy streamers and gas filaments extending inwards from the inner edge of the CND is denoted as the mini-spiral \citep[and references therein]{Lo1983Natur.306..647L, Nitschai2020ApJ...896...68N}. The western arc of the mini-spiral appears to have a circular orbit similar to the neutral gas in the CND. However, the northern and the eastern arms penetrate deep into the ionized cavity on eccentric orbits and reach up to a few arcseconds from Sgr~A*, possibly colliding in the Bar region \citep{Becklin1982ApJ...258..135B, Jackson1993ApJ...402..173J,  Christopher2005ApJ...622..346C, 2009ApJ...699..186Z}. 

There have been several studies on the kinematics of the mini-spiral. \citet{Vollmer2000NewA....4..581V} derived a 3D kinematic model of gas streams that describes the bulk motion of the mini-spiral in three different planes based on the data cube of [NeII] line ($12.8\,{\rm \mu m}$), with the main plane coinciding with the inner rim of the CND. \citet{Paumard2006ApJ...643.1011P} performed a  kinematic study and showed that the Northern arm consists of a weak continuous surface that is drawn into a narrow stream near Sgr~A*. Using $L'$-band data (3.8 ${\rm \mu m}$), \citet{Muzic2007A&A...469..993M} provided proper motions of a number of thin dusty filaments along the mini-spiral and considered a central, partially collimated outflow as the possible explanation for their formation and motion along the mini-spiral, with some deviations from a purely Keplerian rotation. Based on the radio observations using the H92$\alpha$ and H30$\alpha$ lines, \citet{2009ApJ...699..186Z} and \citet{2010ApJ...723.1097Z} determined the 3D velocity field of the mini-spiral. The ionized streamers can dynamically be modeled as a system of three bundles of quasi-Keplerian orbits in the potential dominated by the central mass of $\lesssim 10^7\,M_{\odot}$,  i.e. dominated by Sgr~A* \citep{2009ApJ...699..186Z,2010ApJ...723.1097Z}. The orbital planes of the Northern Arm and the Western Arc are nearly coplanar, while the Eastern Arm plane is perpendicular to them. 

In the central parsec, there is a nuclear stellar cluster (NSC) that consists predominantly of the nearly spherical old cluster of late-type stars as well as the cusp of $\sim 100$ massive young OB/Wolf-Rayet stars that supply about $3 \times 10^{-3} M_{\odot} {\rm yr}^{-1}$ in the form of stellar winds \citep{Najarro1997A&A...325..700N, Moultaka2004A&A...425..529M,2014CQGra..31x4007S}. However, less than 1$\%$  of the Bondi accretion rate is accreted by Sgr~A* \citep{Bower2003ApJ...588..331B, Marrone2006ApJ...640..308M,2013Sci...341..981W}. \cite{Blandford1999MNRAS.303L...1B} proposed a solution for the low accretion rate with the adiabatic inflow-outflow model (ADIOS), in which most of the gas that is accreted has positive energy and is lost through winds and only a small fraction is accreted onto the SMBH. Observational evidence for the outflow first emerged with the detection of the `mini-cavity' region on radio maps by \cite{Yusef-Zadeh1990Natur.348...45Y}. The detection of strong [Fe III] bubble surrounding Sgr~A* \citep{Eckart1992Natur.355..526E,Lutz1993ApJ...418..244L} has led to the conclusion that the fast wind originating within the central few arcseconds blows into the orbiting streamers creating an expanding gas bubble. This has also been supported by \citet{2020A&A...634A..35P} who found that some identified dusty sources located to the west within the S cluster exhibit Doppler-shifted [FeIII] multiplet lines that could be excited by the collimated wind outflowing in their direction towards the mini-cavity. The extended region of low excitation centered on Sgr~A* \citep{Schodel2007A&A...469..125S} and the mass-losing envelope along with the extended tail of IRS7 \citep{Yusef-Zadeh1992ApJ...385L..41Y} could also be influenced by a strong central wind coming from central few arcseconds. Closer to Sgr~A*, the orientation of infrared-excess comet-shaped sources X7, X3, and X8 suggests the presence of a fast, collimated outflow that originates in Sgr~A* accretion flow or in the collective wind of OB/WR stars \citep{2010A&A...521A..13M,2019A&A...624A..97P,2020MNRAS.499.3909Y,Peissker2021ApJ...909...62P}. The nuclear outflow that nearly balances the inflow is also consistent with the flat number density profile of hot plasma, $n_{\rm e}\propto r^{-0.5}$ inside the Bondi radius, as inferred from the analysis of the X-ray bremsstrahlung surface-density profile \citep{2013Sci...341..981W}. In the broader context, the past active jet of Sgr~A* could have contributed to the depletion of bright red giants in the Galactic center due to the intense stripping and the truncation of their extended envelopes \citep{2020ApJ...903..140Z,2020arXiv201112868Z,2021bhns.confE...1K}. 

To fully understand the dynamics of these processes, it becomes important to get proper motions {or tangential velocities} of the various parts of CND as well as the mini-spiral. The dynamics as well as the physical properties of gaseous-dusty structures within the sphere of the gravitational influence of Sgr~A* are crucial for the understanding of the mass and the momentum transport from larger to smaller scales all the way to Sgr~A*.

In this work, we analyze the mid-infrared (MIR) images of the central parsec at two wavelengths in the $N$-band -- $8.59$ (PAH1 filter) and $13.04\,{\rm \mu m}$ (NeII\_2 filter) -- over the course of 12 years. This allows us to study the proper motions of infrared-excess sources of the NSC as well as of identified extended objects in the mini-spiral region. The photometric information at two wavelengths enables us to infer the spectral indices that shed light on the properties of the identified sources. We manage to identify mid-infrared components associated with the circumstellar material of two late-type stars, IRS7 and IRS3, which manifest their interaction with the circumnuclear medium.

The paper is structured as follows. In Section~\ref{sec:DataObservations}, we describe the used dataset and the imaging tools that were applied. Subsequently, in Section~\ref{sec:Results}, we describe the main results of the analysis, including the MIR differential map, tangential velocities, and the photometry, including spectral indices. In particular, we focus on the identification of MIR components of IRS7, the extended circumstellar structure of IRS3, and the general characteristics of the identified dusty sources. We discuss the results in Section~\ref{sec:Discussion}, and subsequently conclude with Section~\ref{sec:Summary}.

\section{Data \& Observations}
\label{sec:DataObservations}

\subsection{VISIR}
\label{sub:VISIR}

\begin{table}[b]
\caption{Details of the observed VISIR images at each epoch.}
\label{tab:Epochs}
\centering
\begin{tabular}{c c c}
\hline
\hline
Year & Pixel Scale & FOV of a single image \\                
\hline
2006 & 0".075 & 19.2" x 19.2" \\
2007 & 0".075 & 19.2" x 19.2" \\
2010 & 0".127 & 32.5" x 32.5"\\
2016 & 0".045 & 38.0" x 38.0" \\
2018 &  0".045 & 38.0" x 38.0"\\
\hline
\end{tabular}
\end{table}

\begin{figure*}
\centering 
\begin{center}
\subfloat{
\includegraphics[width=0.8\textwidth, trim = 0 45 0 0, clip]{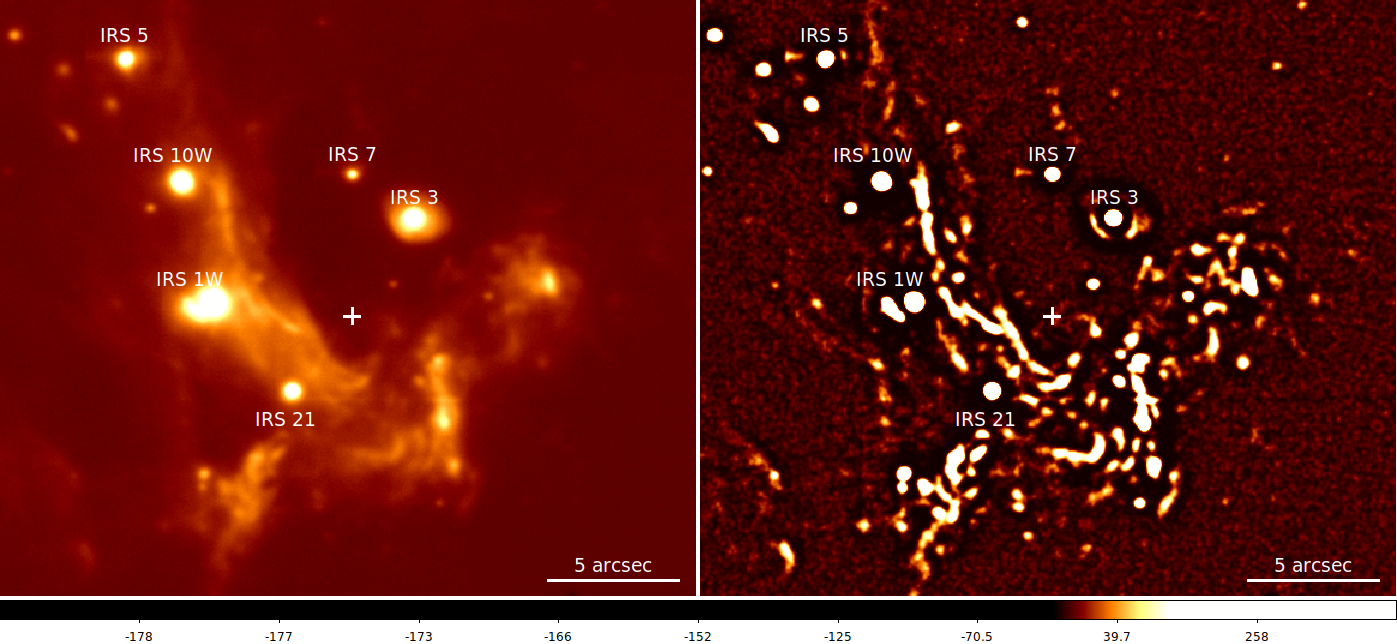}
}
\quad
\subfloat{
\includegraphics[width=0.8\textwidth, trim = 0 45 1 15, clip]{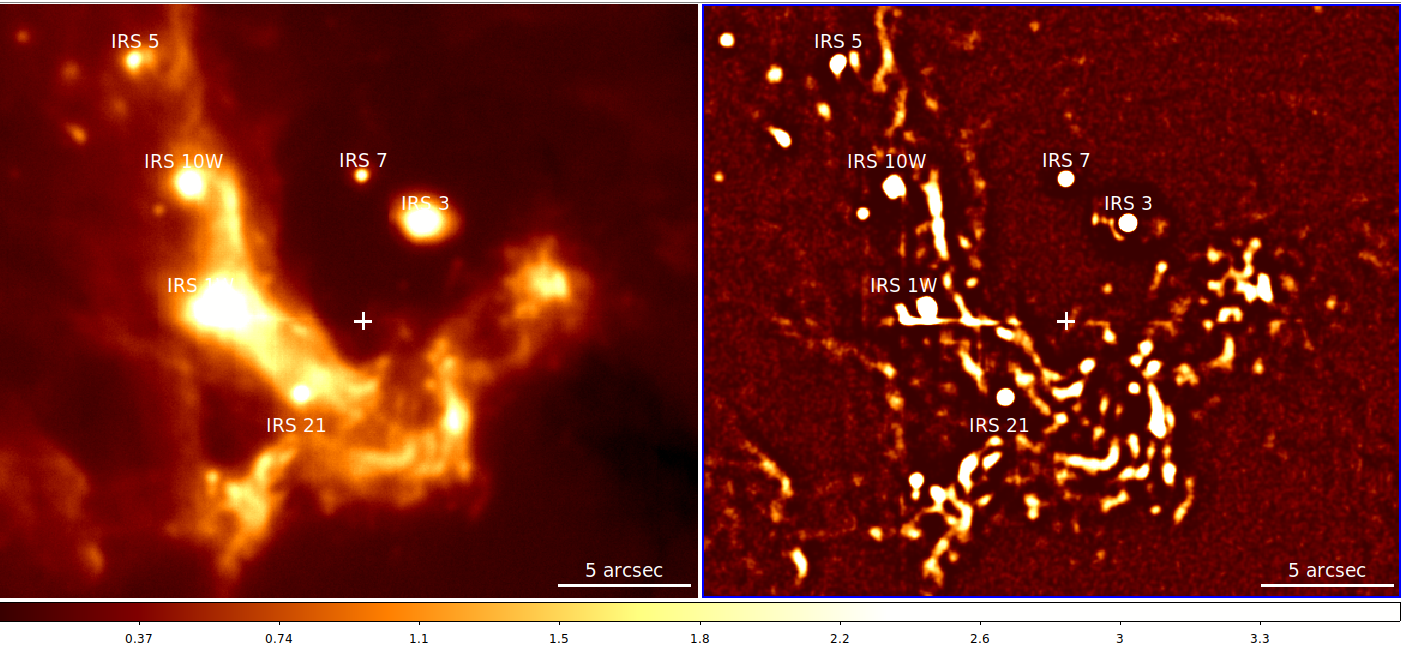}
}
\caption{Central ~24" $\times$ 24" region of the Milky Way in PAH1 filter ($8.59\,{\rm \mu m}$) in the top panel and and in NeII\_2 filter ($13.04\,{\rm \mu m}$) in the bottom panel. The north is up and the east is to the left. The right panel displays the high-pass filtered version of the left image. All the dusty filament-like structures and Infrared-excess stars are resolved and enhanced through the high-pass filter. }
\label{fig:Highpass}
\end{center}
\end{figure*}

Observations were carried out at ESO VLT (UT3) using the VLT Imager and Spectrometer for the mid-InfraRed (VISIR) at 5 different epochs covering a total time from 2006 to 2018 in the N-band PAH1 (8.59$\mu$m) and NeII\_2 (13.04$\mu$m). {Depending on the atmospheric conditions, VISIR provides 0.25”- 0.4” angular resolution imaging}  with high sensitivity. The data used here was obtained as a part of a larger survey (Sabha \etal in prep). The field of view (FOV) and spatial pixel size at each epoch is listed in Table \ref{tab:Epochs}. To increase the signal-to-noise ratio (S/N), jittered images with different offsets were added to create a mosaic with a slightly larger Field Of View. To reduce the bright and varying MIR background, differential observations with the chop/nod mode were executed.  

\subsection{High-pass filter}
\label{sub:High-Pass}

MIR images generally show dusty structures, dust-embedded sources, stellar sources, and even emission from the mini-spiral. Overlapping wings of point-spread functions (PSFs) can create artificial sources and together with noise can complicate the identification of extended objects of our interest, especially in a crowded field of view as the central few arc-seconds of GC. In order to obtain high-angular resolution information and to highlight the structures of the extended sources, we produce high-pass filter maps. High-pass filters can be used as a sharpener and to resolve the objects that are close to the detection limit while preserving the shape of the extended objects. This filtering technique and its significance are described in detail in \citet{Muzic2007A&A...469..993M}; see also \citet{Peissker2020ApJ...897...28P, Peissker2021ApJ...909...62P}. 

First, the Gaussian-smoothed (3-9 pixel Gaussian, corresponding to 0.375"-0.405") version of the input image is subtracted from itself. After removing the negatives, the resulted image is smoothed again using a Gaussian whose size is adjusted depending on the required angular resolution and sensitivity. In Fig.~\ref{fig:Highpass}, we show the effectiveness of this technique for isolating the signals and thereby reducing the change of confusion between nearby objects.


\section{Results}
\label{sec:Results}

\subsection{Differential Map}
\label{sub:DifferentialMap}

\begin{figure}
\begin{center}
\includegraphics[width=0.47\textwidth, trim = 5 3 0 0, clip]{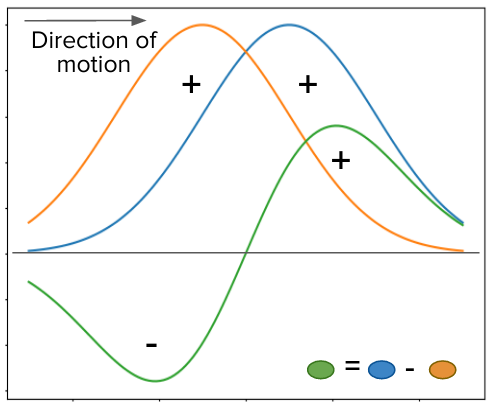}
\caption{Difference of shifted Gaussian profiles as an indicator of the direction of motion.}
\label{fig:Difmap1}
\end{center}
\end{figure}
\begin{figure}
\begin{center}
\includegraphics[width=0.47\textwidth, trim = 0 45 0 0, clip]{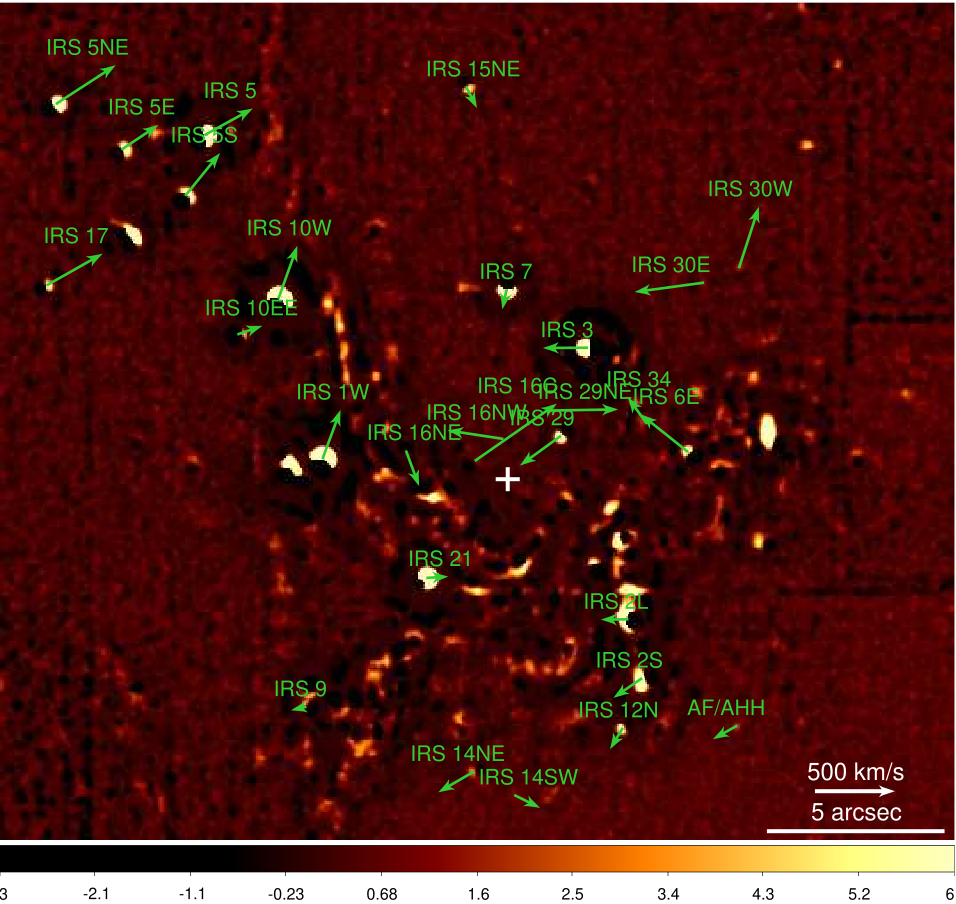}
\caption{Differential map produced by subtracting PAH1 2006 image from PAH1 2018 image. The arrows correspond to velocities obtained by measuring offset positions and velocities w.r.t.  IRS10EE, 9 and 7.}
\label{fig:Difmap2}
\end{center}
\end{figure}

 To get a general idea of the direction of the bulk motion, we produce a differential map using the {smooth-subtracted} images from 2006 and 2018 epochs. As illustrated in Fig. \ref{fig:Difmap1}, subtracting the images separated by large time scales essentially represents the subtraction of two slightly shifted Gaussian profiles. The result is a crude but clear indication of the direction of the bulk motion. { {We first scaled (re-binned) and transform all PAH1 image frames into the common coordinate system of the 2018 epoch using the positions of IRS10EE, IRS9, and IRS7 (and NeII\_2 image frames using the positions of IRS10EE, IRS7, IRS12N, IRS15NE, and IRS17)  to calculate the transformation matrix. These positions were corrected for stellar {velocities} as reported in $K$-band by \cite{Schodel2009} and \cite{Genzel2000MNRAS.317..348G}}}. The resulting image (differential map) is presented in Fig.~\ref{fig:Difmap2} along with the derived proper motions of stellar sources which are discussed in the subsequent subsections. 

\subsection{Proper Motions}
\label{sec:ProperMotion}

\begin{figure*}
\begin{center}
\includegraphics[width=0.99\textwidth, trim = 0 2 2 4, clip]{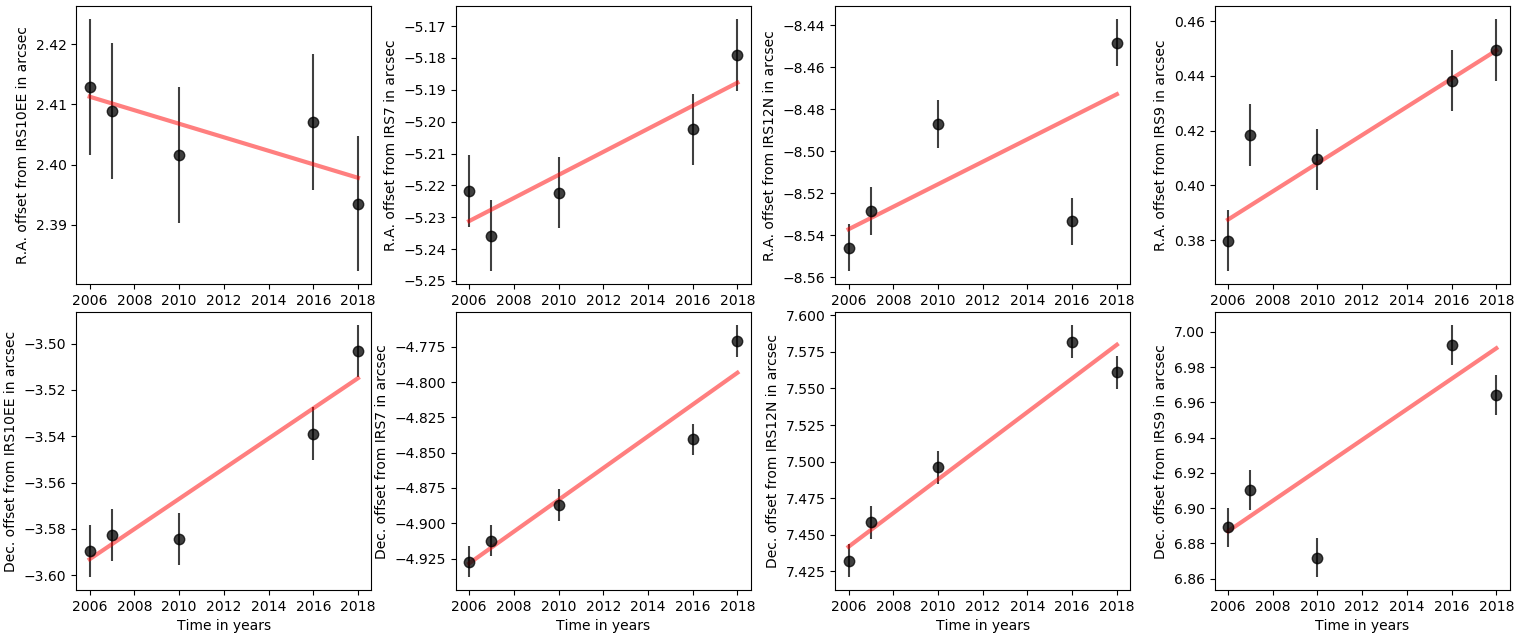}
\caption{Proper motion $v_{\alpha}$ and $v_{\delta}$ of an example source `IRS1W'. Relative velocity w.r.t. IRS10EE, 7, 12N and 9 was determined using offset measurements. Error bars correspond to 0.25 pixels.}
\label{fig:RA_Dec}
\end{center}
\end{figure*}

\begin{figure*}
\begin{center}
\includegraphics[width=0.99\textwidth, trim = 0 45 0 0, clip]{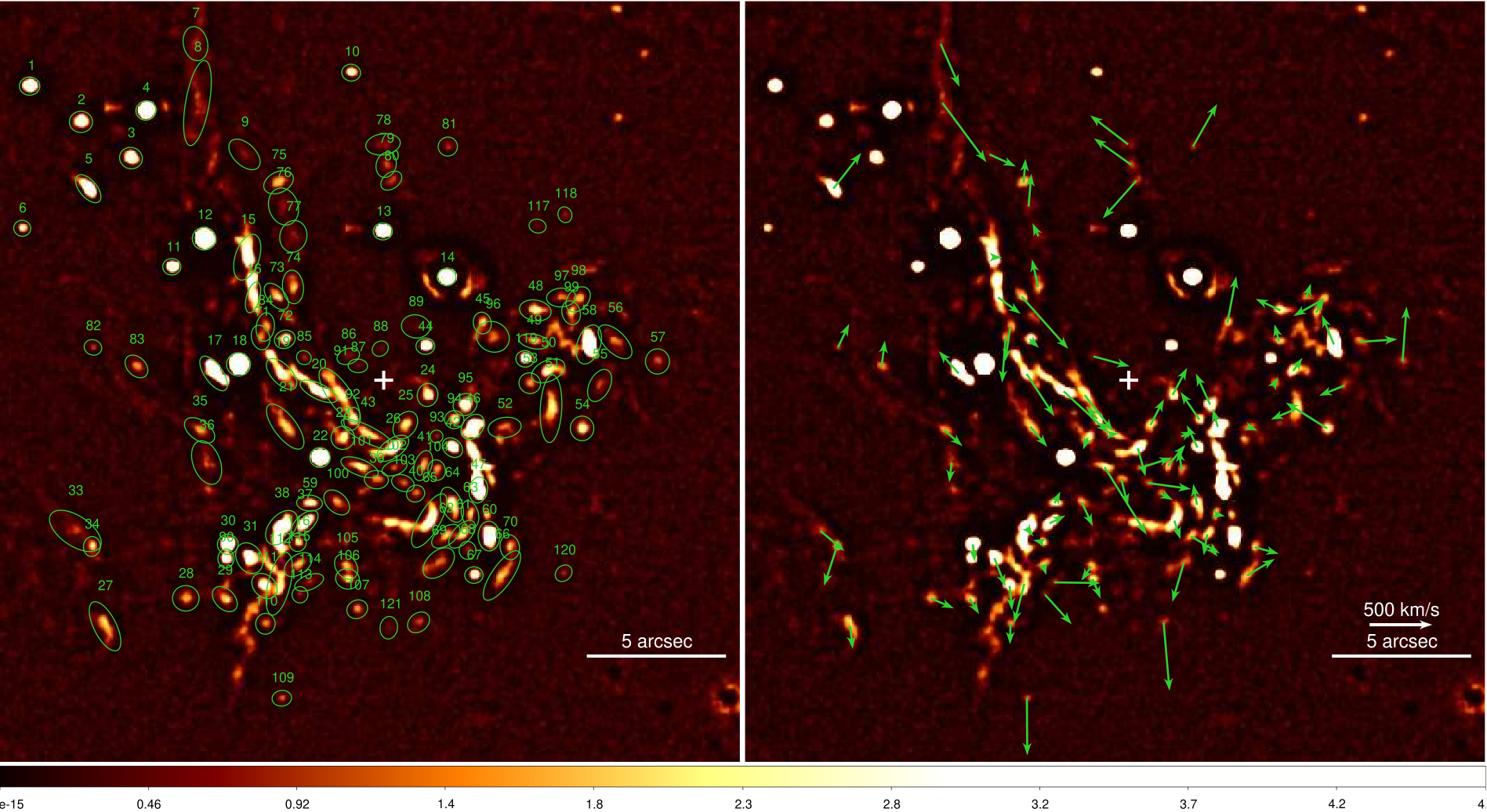}
\caption{The {left} panel is the PAH1 image  with the identification for each source and the apertures used to perform photometry. The {right} panel has proper motion of all the labelled sources. Arrows corresponding to IRS sources are removed for a better visualisation of the stream-like motion.}
\label{fig:Propermotions1}
\end{center}
\end{figure*}

\begin{figure}
\begin{center}
\includegraphics[width=0.47\textwidth, trim = 2600 180 520 300, clip]{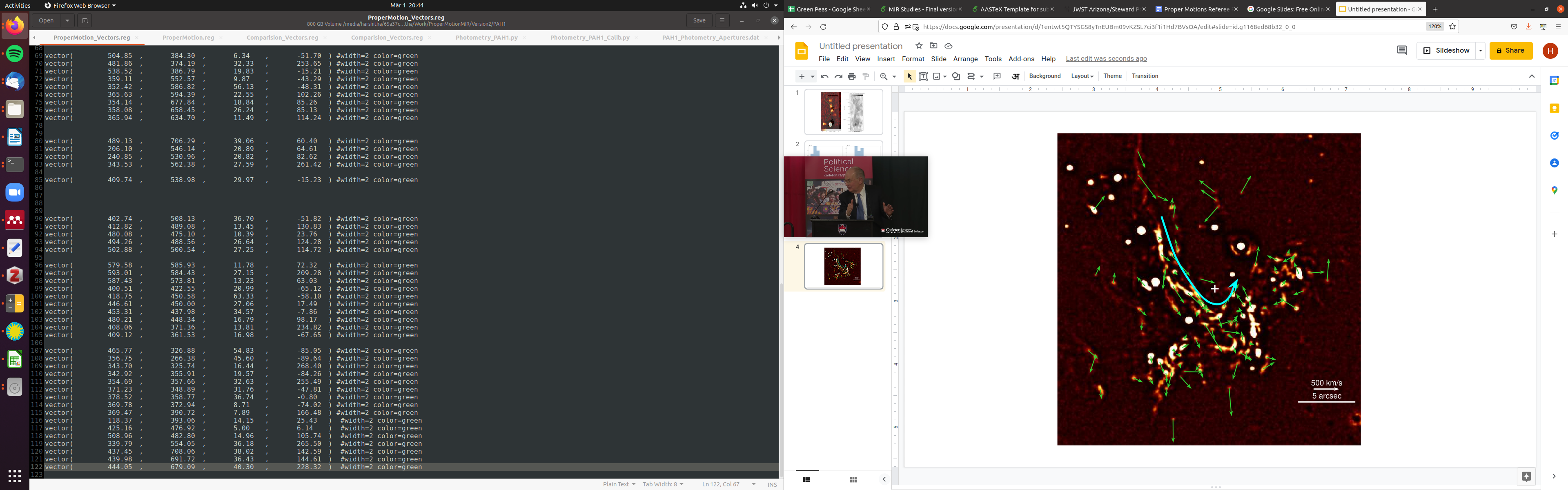}
\caption{Presence of a stream-like motion along the inner edge of the Northern arm towards the S-W direction which changes direction sharply as it crosses the Sgr A* to move in the N-W direction (blue-dashed arrow).}
\label{fig:Propermotions2}
\end{center}
\end{figure}

\begin{table*}[h]
\renewcommand\thetable{2}
\caption{List of tangential velocities. The positions are offsets from SgA* (in 2018) in arcsecs and all the velocities are in ${\rm km\,s^{-1}}$. We choose a conservative 0.25 pixel uncertainties for proper motion calculations which correspond to about 45$ km/s$.}
\label{tab:Propermotions1}
\centering
\begin{tabular}{r c   r r   r r   r r  }
\hline
\hline
&&&&\multicolumn{2}{c}{PAH1} &\multicolumn{2}{c}{NeII\_2}\\
Source & Name & $\Delta \alpha$ & $\Delta \delta$ &  $ v_{\alpha}$ &  $v_{\delta}$ & $ v_{\alpha}$ &  $v_{\delta}$\\

& & (arcsec) & (arcsec) &  \multicolumn{4}{c}{all velocities are in ${\rm km\,s^{-1}}$}  \\

\hline	
1	&	IRS 5NE	&	12.71	&	10.55	&	-374	&			243	&			-167	&			-167			\\
2	&	IRS 5E	&	10.87	&	9.27	&	-232	&			155	&			-216	&			-211			\\
3	&	IRS 5S	&	9.07	&	7.97	&	-218	&			268	&			-75	&			97			\\
4	&	IRS 5	&	8.53	&	9.67	&	-303	&			167	&			-71	&			-145			\\
5	&		&	10.60	&	6.87	&	-223	&			286	&			-226	&			1			\\
6	&	IRS 17	&	12.97	&	5.45	&	-351	&			198	&			-238	&			-83			\\
7	&		&	6.75	&	12.08	&	-151	&			-345	&			-187	&			-279			\\
8	&		&	6.68	&	9.95	&	-350	&			-472	&			-55	&			-52			\\
9	&		&	5.00	&	8.10	&	-210	&			-96	&				&						\\
10	&	IRS 15NE	&	1.17	&	11.04	&	-67	&			-133	&			-6	&			-308			\\
11	&	IRS 10EE	&	7.60	&	4.07	&	-157	&			54	&			-186	&			142			\\
12	&	IRS 10W	&	6.45	&	5.08	&	-118	&			331	&			120	&			148			\\
13	&	IRS 7	&	0.03	&	5.34	&	27	&			-125	&			232	&			-250			\\
14	&	IRS 3	&	-2.26	&	3.70	&	282	&			-2	&			412	&			-13			\\
15	&		&	4.90	&	4.41	&	-77	&			-3	&			-7	&			-182			\\
16	&		&	4.69	&	2.97	&	-180	&			-132	&			-215	&			-215			\\
17	&		&	6.10	&	0.24	&	159	&			178	&			20	&			-51			\\
18	&	IRS 1W	&	5.21	&	0.57	&	-113	&			307	&			53	&			275			\\
19	&		&	3.67	&	0.22	&	-221	&			-337	&			301	&			-218			\\
20	&		&	2.34	&	-0.40	&	-360	&			-330	&			163	&			-205			\\
21	&		&	3.54	&	-1.67	&	-136	&			-162	&			-25	&			-303			\\
22	&	IRS 21	&	2.29	&	-2.78	&	-137	&			8	&			-65	&			3			\\
23	&		&	1.47	&	-2.07	&	38	&			-93	&			196	&			17			\\
24	&		&	-1.57	&	-0.53	&	-116	&			218	&			-170	&			39			\\
25	&		&	-0.78	&	-1.61	&	-108	&			207	&			-180	&			255			\\
26	&		&	-0.33	&	-2.44	&	-61	&			-225	&			-395	&			-33			\\
27	&		&	10.00	&	-8.83	&	-21	&			-181	&			-62	&			-159			\\
28	&		&	7.10	&	-7.83	&	-184	&			-80	&			173	&			-277			\\
29	&		&	5.70	&	-7.86	&	-74	&			-120	&			-259	&			-455			\\
30	&		&	5.60	&	-5.89	&	-16	&			-138	&			63	&			-175			\\
31	&		&	4.82	&	-6.39	&	-89	&			-228	&			316	&			37			\\
32	&		&	14.38	&	-5.71	&	-128	&		61		&			-72	&			-161			\\
33	&		&	11.07	&	-5.41	&	-191	&			-160	&			-44	&			-267			\\
34	&		&	10.47	&	-5.97	&	103	&			-313	&			-135	&			-7			\\
35	&		&	6.61	&	-1.80	&	-142	&			-126	&			-141	&			-195			\\
36	&		&	6.35	&	-2.96	&	20	&			-157	&			64	&			-193			\\
37	&		&	2.84	&	-5.14	&	-126	&			63	&			-185	&			7			\\
38	&		&	3.66	&	-5.29	&	-53	&			-70	&			55	&			-3			\\
39	&		&	0.26	&	-3.57	&	-51	&			-209	&			-189	&			-198			\\
40	&		&	-1.14	&	-4.07	&	-98	&			-52	&			-57	&			-353			\\
41	&		&	-1.41	&	-3.06	&	-139	&			166	&			6	&			-209			\\
42	&		&	-2.47	&	-2.42	&	86	&			180	&			76	&			175			\\
43	&		&	0.58	&	-1.94	&	-50	&	5			&			-295	&			-7			\\
44	&	IRS 29	&	-1.50	&	1.22	&	256	&			-186	&			585	&			-205			\\
45	&		&	-3.54	&	2.05	&	-71	&			374	&			-89	&			502			\\

\hline
\end{tabular}
\end{table*}

\begin{table*}[]
\renewcommand\thetable{2}
\centering
\caption{continued.}
\label{tab:Propermotions}
\begin{tabular}{r c   r r   r r   r r  }
\hline
\hline
&&&&\multicolumn{2}{c}{PAH1} &\multicolumn{2}{c}{NeII\_2}\\
Source & Name & $\Delta \alpha$ & $\Delta \delta$ &  $ v_{\alpha}$ &  $v_{\delta}$ & $ v_{\alpha}$ &  $v_{\delta}$\\

& & (arcsec) & (arcsec) &  \multicolumn{4}{c}{all velocities are in ${\rm km\,s^{-1}}$}  \\

\hline
46	&		&	-3.19	&	-1.67	&	41  &	144			&			93	&			122			\\
47	&	IRS 2L	&	-3.39	&	-3.94	&	173	&			-2	&			293	&			191			\\
48	&		&	-5.44	&	2.50	&	198	&			92	&			362	&			74			\\
49	&		&	-5.39	&	1.39	&	36	&			138	&			699	&			92			\\
50	&		&	-5.90	&	0.33	&	-74	&			104	&			418	&			92			\\
51	&		&	-6.00	&	-1.03	&	202	&			-87	&			100	&			185			\\
52	&		&	-4.34	&	-1.70	&	-61	&			-21	&			185	&			180			\\
53	&		&	-5.24	&	-0.11	&	-27	&			45	&			546	&			-85			\\
54	&		&	-7.12	&	-1.73	&	329	&			206	&			368	&			-79			\\
55	&		&	-7.75	&	-0.19	&	203	&			-81	&			273	&			158			\\
56	&		&	-8.30	&	1.38	&	-297	&			20	&			98	&			181			\\
57	&		&	-9.82	&	0.67	&	-32	&			438	&			123	&			179			\\
58	&		&	-7.36	&	1.28	&	96	&			234	&			15	&			160			\\
59	&		&	2.66	&	-4.41	&	-46	&			93	&			-84	&			357			\\
60	&	IRS 2S	&	-3.78	&	-5.59	&	180	&			-124	&			82	&			-230			\\
61	&		&	-2.81	&	-5.53	&	27	&			-168	&			-255	&			-15			\\
62	&		&	-2.23	&	-5.60	&	-230	&			-137	&			168	&			-139			\\
63	&		&	-3.10	&	-4.81	&	-79	&			-13	&			331	&			36			\\
64	&		&	-2.45	&	-4.47	&	30	&			166	&			-71	&			287			\\
65	&		&	-1.63	&	-5.04	&	-49	&			-149	&			-366	&			145			\\
66	&		&	-4.24	&	-7.01	&	-222	&			132	&			-100	&			-24			\\
67	&	IRS 12N	&	-3.24	&	-6.99	&	80	&			-133	&			-5	&			-106			\\
68	&		&	-3.01	&	-6.11	&	-39	&			-50	&			-127	&			123			\\
69	&		&	-1.97	&	-6.56	&	91	&			-310	&			-326	&			-66			\\
70	&		&	-4.52	&	-5.99	&	-191	&			-52	&			1	&			343			\\
71	&		&	4.42	&	1.53	&	351	&	259			&			132	&			-271			\\
72	&		&	3.55	&	1.47	&	-72	&			-68	&			-150	&			-244			\\
73	&		&	3.85	&	3.01	&	-373	&			-419	&			334	&			-96			\\
74	&		&	3.26	&	3.35	&	48	&			220	&			-21	&			-593			\\
75	&		&	3.77	&	7.10	&	-16	&			188	&				&						\\
76	&		&	3.60	&	6.23	&	-22	&			261	&			-146	&			-338			\\
77	&		&	3.24	&	5.16	&	47	&			105	&			108	&			173			\\
78	&		&	0.02	&	8.46	&	302	&	231			&				&						\\
79	&		&	-0.09	&	7.73	&	297	&	211			&			516	&			126			\\
80	&		&	-0.27	&	7.16	&	268	&	-301			&			990	&			250			\\

81	&		&	-2.30	&	8.38	&	-193	&			340	&			970	&			-529			\\
82	&		&	10.44	&	1.18	&	-90	&			189	&				&						\\
83	&		&	8.87	&	0.49	&	-27	&			207	&				&						\\
84	&		&	4.25	&	1.91	&	41	&			-273	&				&						\\
85	&	IRS 16NE	&	2.87	&	0.81	&	-83	&			-224	&				&						\\
86	&		&	1.27	&	0.85	&	-289	&			-79	&				&						\\
87	&	IRS 16C	&	0.93	&	0.51	&	-513	&			358	&				&						\\
88	&	IRS 16NW	&	0.12	&	1.13	&	351	&			54	&				&						\\
89	&	IRS 29NE	&	-1.15	&	1.93	&	-435	&			7	&				&						\\
90	&	IRS 9	&	5.66	&	-6.40	&	97	&			-22	&				&						\\

\hline
\end{tabular}
\end{table*}

\begin{table*}[]
\renewcommand\thetable{2}
\caption{continued.}
\label{tab:Propermotions}
\centering
\begin{tabular}{r c   r r   r r   r r  }
\hline
\hline
&&&&\multicolumn{2}{c}{PAH1} &\multicolumn{2}{c}{NeII\_2}\\
Source & Name & $\Delta \alpha$ & $\Delta \delta$ &  $ v_{\alpha}$ &  $v_{\delta}$ & $ v_{\alpha}$ &  $v_{\delta}$\\

& & (arcsec) & (arcsec) &  \multicolumn{4}{c}{all velocities are in ${\rm km\,s^{-1}}$}  \\

\hline
91	&		&	1.59	&	-0.53	&	-227	&			-289	&				&						\\
92	&		&	1.13	&	-1.39	&	88	&			102	&				&						\\
93	&		&	-1.89	&	-2.02	&	-95	&			42	&				&						\\
94	&		&	-2.53	&	-1.41	&	150	&			220	&				&						\\
95	&		&	-2.92	&	-0.88	&	114	&			247	&				&						\\
96	&	IRS 34	&	-3.91	&	1.56	&	119	&			166	&				&						\\
97	&		&	-6.37	&	2.97	&	-36	&			112	&				&						\\
98	&		&	-6.98	&	2.90	&	237	&			-133	&				&						\\
99	&		&	-6.72	&	2.42	&	-60	&			118	&				&						\\
100	&		&	1.69	&	-4.39	&	-88	&			-190	&			-100	&			-91			\\
101	&		&	0.87	&	-3.12	&	-335	&			-538	&				&						\\
102	&		&	-0.39	&	-3.15	&	-258	&			81	&			-105	&			-43			\\
103	&		&	-0.69	&	-3.69	&	-342	&			-47	&				&						\\
104	&		&	-1.90	&	-3.22	&	24	&			166	&			-205	&			13			\\
105	&		&	1.35	&	-6.69	&	80	&			-113	&			112	&			100			\\
106	&		&	1.30	&	-7.13	&	-65	&			-157	&				&						\\
107	&	IRS 14NE	&	0.95	&	-8.22	&	223	&			-130	&			153	&			-91			\\
108	&		&	-1.25	&	-8.69	&	-47	&			-546	&				&						\\
109	&		&	3.66	&	-11.41	&	-3	&			-456	&				&						\\
110	&		&	4.24	&	-8.74	&	5	&			-164	&				&						\\
111	&		&	4.28	&	-7.38	&	-20	&			-195	&				&						\\
112	&		&	3.75	&	-7.31	&	82	&			-316	&				&						\\
113	&		&	3.00	&	-7.70	&	-213	&			-235	&				&						\\
114	&		&	2.68	&	-7.26	&	-367	&			-5	&				&						\\
115	&		&	3.07	&	-6.62	&	-24	&			-84	&			-458	&			338			\\
116	&		&	3.08	&	-5.82	&	77	&			18	&				&						\\
117	&	IRS 30E	&	-5.52	&	5.53	&	435	&			-59	&				&						\\
118	&	IRS 30W	&	-6.50	&	5.94	&	-125	&			384	&				&						\\
119	&	IRS 6E	&	-5.07	&	0.76	&	294	&			234	&			398	&			181			\\
120	&	AF/AHH	&	-6.46	&	-6.93	&	152	&			-85	&				&						\\
121	&	IRS 14SW	&	-0.18	&	-8.88	&	-162	&			-80	&				&						\\

\hline
\end{tabular}
\end{table*}
We first identified the IRS sources in N-band images by comparing them with K and L'-band. To make sure the proper motions that we calculate are reliable we make similar calculations for IRS sources in K-band and compare them both with the velocities reported by {\cite{Schodel2009} and  \citet{Genzel2000MNRAS.317..348G}. We choose IRS10EE, IRS9, IRS12N and IRS7 as calibrators for PAH1 images (and IRS10EE, IRS7, IRS12N, IRS15NE, and IRS17 for NeII\_2 images) as they were unambiguously identifiable in both N and K-bands and had velocities $v_{\alpha}$ and $v_{\delta}$ from \citet{Schodel2009}.} We measure the offsets of each source of interest from these IRS sources. We calculate relative velocities w.r.t. each of them and then add their respective velocities (as given by \citeauthor{Schodel2009}, \citeyear{Schodel2009}), before averaging them to get the proper motions w.r.t. Sgr~A*. By doing so, we are minimizing the errors introduced during the process of offset measurements. As an example, we show in Fig. \ref{fig:RA_Dec} the $v_{\alpha}$ and $v_{\delta}$ plots for the source IRS1W. {Multpilying the proper motions with the distance to the Galactc center $D_{GC}=8.1 kpc$ gives the tangential velocity.} We compare {tangential velocities} in N-band and K-band with \cite{Schodel2009} and  \citet{Genzel2000MNRAS.317..348G} in Appendix \ref{sec:ProperMotion_Comp}. 
Table \ref{tab:Comparision_pm} in Appendix \ref{sec:ProperMotion_Comp} contains the calculated velocities in various bands along with the literature values. Fig. \ref{fig:Comparision_pm} and Fig. \ref{fig:Deviation}  in Appendix \ref{sec:ProperMotion_Comp} depict the deviations from \cite{Schodel2009}. {It is important to note that the uncertainties in the fainter sources could be larger than in the bright sources due to the possibility of unresolved background sources blending with our target sources.} {While the uncertainties of the Gaussian fits for the positions is typically of the order of a few hundredth of a pixel we conservatively assume an uncertainty of 0.25 pixels for that quantity}.  {The mean absolute difference between the velocities in 2 filters is about 100 km/s and standard deviation of the angle difference is about 40 degrees}.

We then move on to calculate the tangential velocities of all the extended sources in the FOV, concentrating especially on the inner edge of the Northern arm. We determine the positions of each resolved source by fitting an elliptical Gaussian using the data visualization tool, \texttt{QFitsView} \citep{Ott2012ascl.soft10019O}. We derive tangential velocities as described above and tabulate them in Table \ref{tab:Propermotions1}. We only report velocities of those sources that are free from confusion in {at least} 3 epochs. The uncertainties in the combined R.A./Dec velocities are about $\pm$ 120 km/s and the uncertainties in the flight direction are about $\pm$ 30 degrees. Fig \ref{fig:Propermotions1} shows the source labels and the used apertures in the {\it{Left panel}} and their derived proper motion vectors in the {\it{Right panel}}. We have removed the velocity vectors belonging to the IRS sources to get a clear indication of any bulk motion. We see a clear stream-like motion along the inner edge of the Northern arm towards the S-W direction which changes direction sharply as it crosses the Sgr~A* to move in the N-W direction (blue dashed arrow in Fig. \ref{fig:Propermotions2}). 

{{
We assume that within the area of the mini-spiral Northern Arm  we find $N_{NA}$ sources that truly belong to the Northern Arm. The number of source in that region that belong to the 
underlying cluster is $N_{C}$. We assume that $N_C$ is a fraction of $N_{NA}$, i.e.,

\begin{equation}
N_C = f \times N_{NA}~~~.
\end{equation}

The total number of sources detected within the area of the mini-spiral Northern Arm is then given by

\begin{equation}
N_{tot}=N_C + N_{NA} =  N_{NA} \times (1+f)~~~.
\end{equation}

Therefore, even if all the $N_{NA}$ sources move downstream one would not expect that all sources observed in that region follow that trend. While we assume that all the $N_{NA}$ sources in the Northern Arm will be headed downstream along the mini-spiral arm, we also assume in a very simplistic way that half of the cluster sources $N_C$ in that area will be moving downstream and half of them upstream. This implies that the ratio $R$ between the sources moving up- and downstream is:

\begin{equation}
R = \frac{N_{NA}+0.5~N_{C}}{0.5~N_{C}} =
 \frac{N_{C}/f+0.5~N_{C}}{0.5~N_{C}} =
 \frac{2}{f} +1~~~. 
\end{equation}

As can be seen in Fig. \ref{fig:Propermotions1} we find that in the mini-spiral region 11 sources move downstream and 4 sources move upstream, implying $R=11/4=2.75$. Hence, we find that 

\begin{equation}
R = \frac{2}{f} +1 = 2.75~~~.
\end{equation}

Therefore, that fraction $f$ turns out to be just above unity, with

\begin{equation}
f=\frac{2}{2.75-1}=1.14
\end{equation}

just as expected for an additional contribution of dusty infrared excess sources due to the Northern Arm on top of a cluster contribution with conceivably almost the same number density in dusty, stellar infrared excess sources. For an equal contribution from the Northern Arm and the cluster in the region of the Northern Arm we would have expected $R=3$.
}}
\subsection{Photometry}
\label{sub:Photometry}

\begin{figure}
\begin{center}
\includegraphics[width=0.47\textwidth, trim = 3 2 2 5, clip]{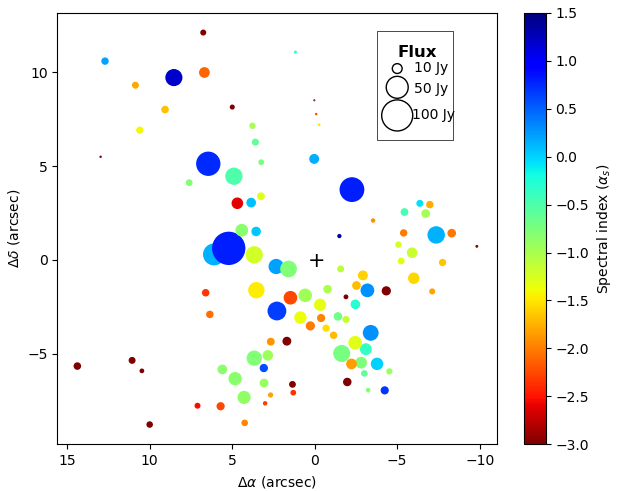}
\caption{Spectral indices using the de-reddened flux density measurements in PAH1 and NeII\_2 filters. The positions are offsets from Sgr~A* and the size corresponds to the flux density in the PAH1 filter. The figure also shows the positive correlation between the PAH1 flux density and the spectral index ($\alpha_{s}$) - the larger the PAH1 flux density (larger circles), the greater the spectral index (bluer colors).}
\label{fig:Spectral_Index}
\end{center}
\end{figure}

\begin{table*}[]
\renewcommand\thetable{3}
\caption{List of reliable flux densities in PAH1 and NeII\_2 filters.}
\label{tab:Photometry}
\centering
\begin{tabular}{r c r r r r r r r r r r r}
\hline
\hline
Source & Name & $F_{8.59}$& $\Delta F_{8.59}$ & $F'_{8.59}$ & $\Delta F'_{8.59}$  & $F_{13.04} $& $\Delta F_{13.04}$ & $F'_{13.04} $& $\Delta F'_{13.04}$ & $ \alpha_{s}$ &  $\Delta \alpha_{s}$ \\

& & (Jy) & (Jy)& (Jy)& (Jy) & (Jy) & (Jy)& (Jy)& (Jy)&&& \\

\hline
1	&	IRS 5NE	&	0.57	&	0.09	&	3.75	&	0.58	&	1	&	0.09	&	3.43	&	0.3	&	0.21	&	0.43	\\
2	&	IRS 5E	&	0.49	&	0.08	&	3.22	&	0.50	&	1.99	&	0.17	&	6.84	&	0.58	&	-1.80	&	0.42	\\
3	&	IRS 5S	&	0.61	&	0.09	&	4.01	&	0.62	&	2.35	&	0.2	&	8.07	&	0.69	&	-1.68	&	0.42	\\
4	&	IRS 5	&	3.92	&	0.60	&	25.64	&	3.90	&	4.54	&	0.38	&	15.58	&	1.32	&	1.19	&	0.42	\\
5	&		&	0.54	&	0.08	&	3.50	&	0.54	&	1.83	&	0.16	&	6.3	&	0.54	&	-1.41	&	0.42	\\
6	&	IRS 17	&	0.02	&	0.01	&	0.11	&	0.02	&	0.27	&	0.02	&	0.94	&	0.09	&	-5.14	&	0.49	\\
7	&		&	0.33	&	0.05	&	2.18	&	0.34	&	3.47	&	0.29	&	11.91	&	1.01	&	-4.07	&	0.43	\\
8	&		&	1.38	&	0.21	&	9.02	&	1.39	&	6.33	&	0.54	&	21.73	&	1.84	&	-2.11	&	0.42	\\
9	&		&	0.21	&	0.03	&	1.38	&	0.22	&	5.36	&	0.46	&	18.43	&	1.56	&	-6.21	&	0.43	\\
10	&	IRS 15NE	&	0.04	&	0.01	&	0.26	&	0.05	&	0.09	&	0.01	&	0.3	&	0.03	&	-0.34	&	0.52	\\
11	&	IRS 10EE	&	0.46	&	0.07	&	3.02	&	0.47	&	1.22	&	0.1	&	4.19	&	0.36	&	-0.78	&	0.43	\\
12	&	IRS 10W	&	8.33	&	1.27	&	54.51	&	8.28	&	11.61	&	0.98	&	39.87	&	3.36	&	0.75	&	0.42	\\
13	&	IRS 7	&	1.15	&	0.18	&	7.54	&	1.15	&	2.04	&	0.17	&	7	&	0.6	&	0.18	&	0.42	\\
14	&	IRS 3	&	8.65	&	1.31	&	56.59	&	8.60	&	11.75	&	0.99	&	40.37	&	3.4	&	0.81	&	0.42	\\
15	&		&	4.00	&	0.61	&	26.17	&	3.99	&	9.39	&	0.79	&	32.26	&	2.72	&	-0.50	&	0.42	\\
16	&		&	1.64	&	0.25	&	10.73	&	1.64	&	9.27	&	0.78	&	31.83	&	2.69	&	-2.60	&	0.42	\\
17	&		&	6.67	&	1.01	&	43.64	&	6.63	&	11.87	&	1	&	40.76	&	3.44	&	0.16	&	0.42	\\
18	&	IRS 1W	&	16.14	&	2.45	&	105.66	&	16.05	&	21.97	&	1.85	&	75.47	&	6.36	&	0.81	&	0.42	\\
19	&		&	3.99	&	0.61	&	26.15	&	3.98	&	12.64	&	1.07	&	43.44	&	3.67	&	-1.22	&	0.42	\\
20	&		&	3.03	&	0.46	&	19.85	&	3.02	&	5.28	&	0.45	&	18.14	&	1.53	&	0.22	&	0.42	\\
21	&		&	3.54	&	0.54	&	23.19	&	3.53	&	12.48	&	1.05	&	42.86	&	3.62	&	-1.47	&	0.42	\\
22	&	IRS 21	&	4.79	&	0.73	&	31.39	&	4.77	&	6.92	&	0.58	&	23.77	&	2.01	&	0.67	&	0.42	\\
23	&		&	2.37	&	0.36	&	15.52	&	2.36	&	11.58	&	0.98	&	39.79	&	3.36	&	-2.26	&	0.42	\\
24	&		&	0.49	&	0.08	&	3.24	&	0.50	&	1.49	&	0.13	&	5.11	&	0.44	&	-1.09	&	0.42	\\
25	&		&	0.81	&	0.12	&	5.28	&	0.81	&	2.34	&	0.2	&	8.02	&	0.68	&	-1.00	&	0.42	\\
26	&		&	1.86	&	0.28	&	12.18	&	1.86	&	6.16	&	0.52	&	21.15	&	1.79	&	-1.32	&	0.42	\\
27	&		&	0.42	&	0.07	&	2.74	&	0.43	&	2.78	&	0.24	&	9.54	&	0.81	&	-2.99	&	0.43	\\
28	&		&	0.33	&	0.05	&	2.13	&	0.33	&	1.78	&	0.15	&	6.12	&	0.52	&	-2.53	&	0.42	\\
29	&		&	0.74	&	0.11	&	4.82	&	0.74	&	3.61	&	0.31	&	12.39	&	1.05	&	-2.26	&	0.42	\\
30	&		&	1.10	&	0.17	&	7.18	&	1.10	&	2.95	&	0.25	&	10.14	&	0.86	&	-0.83	&	0.42	\\
31	&		&	2.26	&	0.35	&	14.81	&	2.26	&	6.09	&	0.52	&	20.92	&	1.77	&	-0.83	&	0.42	\\
32	&		&	0.60	&	0.09	&	3.91	&	0.61	&	4.45	&	0.38	&	15.28	&	1.3	&	-3.27	&	0.43	\\
33	&		&	0.49	&	0.08	&	3.19	&	0.50	&	3.44	&	0.29	&	11.83	&	1.01	&	-3.14	&	0.43	\\
34	&		&	0.18	&	0.03	&	1.15	&	0.18	&	1.89	&	0.16	&	6.48	&	0.55	&	-4.14	&	0.43	\\
35	&		&	0.59	&	0.09	&	3.87	&	0.60	&	2.99	&	0.25	&	10.28	&	0.87	&	-2.34	&	0.42	\\
36	&		&	0.59	&	0.09	&	3.86	&	0.60	&	2.65	&	0.23	&	9.11	&	0.78	&	-2.06	&	0.43	\\
37	&		&	1.32	&	0.20	&	8.65	&	1.32	&	3.72	&	0.32	&	12.79	&	1.08	&	-0.94	&	0.42	\\
38	&		&	3.05	&	0.47	&	19.98	&	3.04	&	8.05	&	0.68	&	27.67	&	2.34	&	-0.78	&	0.42	\\
39	&		&	0.96	&	0.15	&	6.29	&	0.96	&	4.24	&	0.36	&	14.56	&	1.23	&	-2.01	&	0.42	\\
40	&		&	0.58	&	0.09	&	3.78	&	0.58	&	2.25	&	0.19	&	7.72	&	0.66	&	-1.71	&	0.42	\\
41	&		&	0.81	&	0.12	&	5.30	&	0.81	&	2.05	&	0.17	&	7.04	&	0.6	&	-0.68	&	0.42	\\
42	&		&	1.06	&	0.16	&	6.92	&	1.06	&	2.26	&	0.19	&	7.77	&	0.66	&	-0.28	&	0.42	\\
43	&		&	2.33	&	0.36	&	15.24	&	2.33	&	6.6	&	0.56	&	22.68	&	1.92	&	-0.95	&	0.42	\\
44	&	IRS 29	&	0.13	&	0.02	&	0.85	&	0.14	&	0.14	&	0.01	&	0.48	&	0.05	&	1.37	&	0.47	\\
45	&		&	0.14	&	0.02	&	0.94	&	0.15	&	0.61	&	0.05	&	2.09	&	0.18	&	-1.91	&	0.43	\\

\hline
\end{tabular}
\end{table*}

\begin{table*}[]
\renewcommand\thetable{3}
\caption{continued.}
\label{tab:Photometry}
\centering
\begin{tabular}{r c r r r r r r r r r r r}
\hline
\hline
Source & Name & $F_{8.59}$& $\Delta F_{8.59}$ & $F'_{8.59}$ & $\Delta F'_{8.59}$  & $F_{13.04} $& $\Delta F_{13.04}$ & $F'_{13.04} $& $\Delta F'_{13.04}$ & $ \alpha_{s}$ &  $\Delta \alpha_{s}$ \\

& & (Jy) & (Jy)& (Jy)& (Jy) & (Jy) & (Jy)& (Jy)& (Jy)&&& \\

\hline
46	&		&	2.41	&	0.37	&	15.80	&	2.41	&	4.08	&	0.35	&	14	&	1.19	&	0.29	&	0.42	\\
47	&	IRS 2L	&	3.28	&	0.50	&	21.49	&	3.27	&	5.52	&	0.47	&	18.98	&	1.6	&	0.30	&	0.42	\\
48	&		&	0.64	&	0.10	&	4.20	&	0.65	&	1.48	&	0.13	&	5.09	&	0.44	&	-0.46	&	0.42	\\
49	&		&	0.57	&	0.09	&	3.72	&	0.57	&	2.5	&	0.21	&	8.6	&	0.73	&	-2.01	&	0.42	\\
50	&		&	1.33	&	0.20	&	8.71	&	1.33	&	4.11	&	0.35	&	14.14	&	1.2	&	-1.16	&	0.42	\\
51	&		&	1.60	&	0.25	&	10.49	&	1.61	&	5.88	&	0.5	&	20.21	&	1.71	&	-1.57	&	0.42	\\
52	&		&	0.97	&	0.15	&	6.34	&	0.97	&	7.58	&	0.64	&	26.04	&	2.21	&	-3.38	&	0.42	\\
53	&		&	0.46	&	0.07	&	3.01	&	0.46	&	1.51	&	0.13	&	5.17	&	0.44	&	-1.30	&	0.42	\\
54	&		&	0.34	&	0.05	&	2.24	&	0.35	&	1.39	&	0.12	&	4.79	&	0.41	&	-1.82	&	0.43	\\
55	&		&	0.55	&	0.09	&	3.61	&	0.56	&	2.09	&	0.18	&	7.19	&	0.61	&	-1.65	&	0.42	\\
56	&		&	0.82	&	0.13	&	5.38	&	0.83	&	3.65	&	0.31	&	12.55	&	1.07	&	-2.03	&	0.42	\\
57	&		&	0.03	&	0.01	&	0.22	&	0.04	&	0.47	&	0.04	&	1.61	&	0.14	&	-4.77	&	0.48	\\
58	&		&	4.07	&	0.62	&	26.65	&	4.06	&	7.26	&	0.61	&	24.95	&	2.11	&	0.16	&	0.42	\\
59	&		&	0.66	&	0.10	&	4.34	&	0.67	&	2.77	&	0.23	&	9.5	&	0.81	&	-1.88	&	0.42	\\
60	&	IRS 2S	&	1.93	&	0.29	&	12.64	&	1.93	&	3.67	&	0.31	&	12.61	&	1.07	&	0.01	&	0.42	\\
61	&		&	1.68	&	0.26	&	11.00	&	1.68	&	4.4	&	0.37	&	15.1	&	1.28	&	-0.76	&	0.42	\\
62	&		&	1.37	&	0.21	&	8.97	&	1.37	&	5.63	&	0.48	&	19.36	&	1.64	&	-1.84	&	0.42	\\
63	&		&	1.73	&	0.26	&	11.36	&	1.73	&	3.83	&	0.32	&	13.17	&	1.12	&	-0.35	&	0.42	\\
64	&		&	2.34	&	0.36	&	15.34	&	2.34	&	7.67	&	0.65	&	26.37	&	2.23	&	-1.30	&	0.42	\\
65	&		&	3.86	&	0.59	&	25.27	&	3.85	&	9.92	&	0.84	&	34.1	&	2.88	&	-0.72	&	0.42	\\
66	&		&	0.73	&	0.11	&	4.75	&	0.73	&	1.03	&	0.09	&	3.54	&	0.31	&	0.70	&	0.42	\\
67	&	IRS 12N	&	0.16	&	0.03	&	1.05	&	0.17	&	0.42	&	0.04	&	1.46	&	0.13	&	-0.79	&	0.44	\\
68	&		&	0.42	&	0.07	&	2.77	&	0.43	&	1.03	&	0.09	&	3.55	&	0.3	&	-0.59	&	0.42	\\
69	&		&	0.80	&	0.12	&	5.21	&	0.80	&	5.9	&	0.5	&	20.28	&	1.72	&	-3.26	&	0.42	\\
70	&		&	0.36	&	0.06	&	2.36	&	0.37	&	1	&	0.09	&	3.44	&	0.3	&	-0.90	&	0.43	\\
71	&		&	2.02	&	0.31	&	13.20	&	2.01	&	5.43	&	0.46	&	18.64	&	1.58	&	-0.83	&	0.42	\\
72	&		&	1.04	&	0.16	&	6.82	&	1.04	&	1.94	&	0.17	&	6.68	&	0.57	&	0.05	&	0.42	\\
73	&		&	1.08	&	0.17	&	7.09	&	1.08	&	1.98	&	0.17	&	6.82	&	0.58	&	0.09	&	0.42	\\
74	&		&	0.64	&	0.10	&	4.18	&	0.65	&	2.09	&	0.18	&	7.18	&	0.61	&	-1.30	&	0.42	\\
75	&		&	0.39	&	0.06	&	2.56	&	0.40	&	1.12	&	0.1	&	3.85	&	0.33	&	-0.98	&	0.43	\\
76	&		&	0.50	&	0.08	&	3.27	&	0.51	&	1.22	&	0.11	&	4.2	&	0.36	&	-0.60	&	0.43	\\
77	&		&	0.30	&	0.05	&	1.98	&	0.31	&	0.78	&	0.07	&	2.69	&	0.23	&	-0.73	&	0.43	\\
78	&		&	0.01	&	0.01	&	0.05	&	0.02	&	0.08	&	0.01	&	0.27	&	0.03	&	-4.04	&	0.99	\\
79	&		&	0.02	&	0.01	&	0.14	&	0.03	&	0.1	&	0.01	&	0.35	&	0.04	&	-2.20	&	0.58	\\
80	&		&	0.02	&	0.01	&	0.16	&	0.03	&	0.09	&	0.01	&	0.3	&	0.03	&	-1.51	&	0.51	\\
81	&		&		&		&		&		&	0.01	&	0.01	&	0.03	&	0.01	&		&		\\
82	&		&	0.01	&	0.01	&	0.09	&	0.02	&		&		&		&		&		&		\\
83	&		&	0.30	&	0.05	&	1.96	&	0.31	&		&		&		&		&		&		\\
84	&		&	1.69	&	0.26	&	11.05	&	1.69	&		&		&		&		&		&		\\
85	&	IRS 16NE	&	0.50	&	0.08	&	3.26	&	0.50	&		&		&		&		&		&		\\
86	&		&	0.02	&	0.01	&	0.12	&	0.02	&		&		&		&		&		&		\\
87	&	IRS 16C	&	0.01	&	0.01	&	0.05	&	0.01	&		&		&		&		&		&		\\
88	&	IRS 16NW	&		&		&		&		&		&		&		&		&		&		\\
89	&	IRS 29NE	&	0.01	&	0.01	&	0.04	&	0.02	&		&		&		&		&		&		\\
90	&	IRS 9	&	0.78	&	0.12	&	5.14	&	0.79	&		&		&		&		&		&		\\

\hline
\end{tabular}
\end{table*}

\begin{table*}[]
\renewcommand\thetable{3}
\caption{continued.}
\label{tab:Photometry}
\centering
\begin{tabular}{r c r r r r r r r r r r r}
\hline
\hline
Source & Name & $F_{8.59}$& $\Delta F_{8.59}$ & $F'_{8.59}$ & $\Delta F'_{8.59}$  & $F_{13.04} $& $\Delta F_{13.04}$ & $F'_{13.04} $& $\Delta F'_{13.04}$ & $ \alpha_{s}$ &  $\Delta \alpha_{s}$ \\

& & (Jy) & (Jy)& (Jy)& (Jy) & (Jy) & (Jy)& (Jy)& (Jy)&&& \\

\hline
91	&		&	3.73	&	0.57	&	24.39	&	3.72	&	9.83	&	0.83	&	33.77	&	2.86	&	-0.78	&	0.42	\\
92	&		&	1.15	&	0.18	&	7.54	&	1.15	&	&		&	&		&		&		\\
93	&		&	0.19	&	0.03	&	1.22	&	0.19	&	1.24	&	0.11	&	4.27	&	0.37	&	-3.00	&	0.43	\\
94	&		&	0.85	&	0.13	&	5.55	&	0.85	&	3.28	&	0.28	&	11.26	&	0.96	&	-1.69	&	0.42	\\
95	&		&	1.15	&	0.18	&	7.50	&	1.15	&	4.24	&	0.36	&	14.58	&	1.24	&	-1.59	&	0.42	\\
96	&	IRS 34	&	0.44	&	0.07	&	2.86	&	0.45	&		&		&		&		&		&		\\
97	&		&	0.50	&	0.08	&	3.25	&	0.50	&	0.97	&	0.08	&	3.33	&	0.29	&	-0.06	&	0.42	\\
98	&		&	0.57	&	0.09	&	3.74	&	0.58	&	2.29	&	0.2	&	7.87	&	0.67	&	-1.78	&	0.42	\\
99	&		&	0.85	&	0.13	&	5.57	&	0.85	&	2.43	&	0.21	&	8.34	&	0.71	&	-0.97	&	0.42	\\
100	&		&	0.88	&	0.13	&	5.76	&	0.88	&	6.17	&	0.52	&	21.19	&	1.79	&	-3.12	&	0.42	\\
101	&		&	1.94	&	0.30	&	12.67	&	1.93	&	6.48	&	0.55	&	22.26	&	1.88	&	-1.35	&	0.42	\\
102	&		&	0.72	&	0.11	&	4.72	&	0.72	&	3.14	&	0.27	&	10.8	&	0.92	&	-1.98	&	0.42	\\
103	&		&	0.56	&	0.09	&	3.68	&	0.57	&	2.03	&	0.17	&	6.98	&	0.59	&	-1.53	&	0.42	\\
104	&		&	0.52	&	0.08	&	3.42	&	0.53	&	1.61	&	0.14	&	5.52	&	0.47	&	-1.15	&	0.42	\\
105	&		&	0.46	&	0.07	&	3.01	&	0.47	&	3.54	&	0.3	&	12.15	&	1.03	&	-3.34	&	0.43	\\
106	&		&	0.29	&	0.05	&	1.92	&	0.30	&	1.5	&	0.13	&	5.14	&	0.44	&	-2.36	&	0.43	\\
107	&	IRS 14NE	&	0.01	&	0.01	&	0.07	&	0.02	&		&		&		&		&		&		\\
108	&		&		&		&		&		&	0.14	&	0.01	&	0.47	&	0.05	&		&		\\
109	&		&		&		&		&		&		&		&		&		&		&		\\
110	&		&	0.43	&	0.07	&	2.80	&	0.43	&	1.84	&	0.16	&	6.31	&	0.54	&	-1.95	&	0.42	\\
111	&		&	2.18	&	0.33	&	14.24	&	2.17	&	5.94	&	0.5	&	20.41	&	1.73	&	-0.86	&	0.42	\\
112	&		&	3.85	&	0.59	&	25.22	&	3.84	&		&		&		&		&		&		\\
113	&		&	0.15	&	0.02	&	1.00	&	0.16	&	0.76	&	0.07	&	2.6	&	0.22	&	-2.29	&	0.43	\\
114	&		&	0.24	&	0.04	&	1.54	&	0.24	&	0.96	&	0.08	&	3.28	&	0.28	&	-1.81	&	0.43	\\
115	&		&	0.88	&	0.14	&	5.78	&	0.89	&	2.45	&	0.21	&	8.41	&	0.72	&	-0.90	&	0.42	\\
116	&		&	0.76	&	0.12	&	4.96	&	0.76	&	1.13	&	0.1	&	3.87	&	0.33	&	0.59	&	0.42	\\
117	&	IRS 30E	&		&		&		&		&		&		&		&		&		&		\\
118	&	IRS 30W	&		&		&		&		&		&		&		&		&		&		\\
119	&	IRS 6E	&	0.39	&	0.06	&	2.53	&	0.39	&	1.22	&	0.1	&	4.19	&	0.36	&	-1.21	&	0.42	\\
120	&	AF/AHH	&		&		&		&		&		&		&		&		&		&		\\
121	&	IRS 14SW	&		&		&		&		&		&		&		&		&		&		\\

\hline
\end{tabular}
\end{table*}
Even though there have been many photometric studies on stars in the galactic center in various other wavelengths, they are limited in MIR \citep[e.g.][]{Blum1996ApJ...470..864B, Ott1999ApJ...523..248O, Tanner2002, Viehmann2005A&A...433..117V,Viehmann2006ApJ...642..861V}. We carry on these studies on the extended dusty sources that we have identified.

{\cite{Viehmann2006ApJ...642..861V} report flux densities in various filters including N-band. In the MIR regime, they perform relative aperture photometry while relying on \cite{Tanner2002} for flux calibration. Contrary to what is mentioned in \cite{Viehmann2006ApJ...642..861V}, it appears that the reported N-band flux densities are not extinction corrected. For example, \cite{Viehmann2006ApJ...642..861V} report IRS 21 flux density as 4.56 Jy at 8.6$\mu$m; which is consistent with 3.6 Jy by \cite{Stolovy1996} at 8.7$\mu$m and 3.6 by \cite{Tanner2002} at 8.8$\mu$m. \cite{Stolovy1996} results are based on the zero point measurements made with SpectroCam-10 on the 200 inch Hale telescope. As both of these previous studies did not report de-reddened fluxes, if \cite{Viehmann2006ApJ...642..861V} values were in-fact extinction corrected we would expect it to be about 5 times higher. This finding is further confirmed by two other independent studies by Sabha et al. (in prep.) and Schödel et al. (in prep) (through private communication), both based on zero point measurements provided by ESO. }

The flux densities were extracted via the aperture photometry using elliptical apertures to match the shape of extended or filamentary sources (see Fig. \ref{fig:Propermotions1}). Aperture sizes were determined by fitting a 2-D Gaussian along semi-major and semi-minor axes of the {source}. {We selected IRS5NE, IRS10W, IRS7 and IRS1W as calibrators using source aperture and background similar to \cite{Viehmann2006ApJ...642..861V}. As all the elliptical apertures we use do not have the same area, we measured surface density of background contribution at various apertures {in the uncrowded regions} of the image and multiplied it by the area of each ellipse to get their background contributions.} The extinction correction of $A_{\lambda}\sim2.04$ for PAH1 and $A_{\lambda}\sim 1.34$ for Ne II \citep{Fritz2011ApJ...737...73F} was applied to obtain de-reddened flux densities.

{Table \ref{tab:Flux_Comparions} in Appendix \ref{sec:Flux_Comparison} lists the flux density values by \cite{Viehmann2006ApJ...642..861V}, our results when we use similar apertures and background as \cite{Viehmann2006ApJ...642..861V}, and our results when we use aperture sizes determined by FWHM of the 2D Gaussian that we fit (our results in the table are before extinction correction). The table shows that our calibration approach results in source flux densities that are in good agreement with the results from \cite{Viehmann2006ApJ...642..861V}. It is to be noted that the central wavelength of the NeII\_2 filter used for our images is 13.04 $\mu m$ (NeII\_2), while that of \cite{Viehmann2006ApJ...642..861V} images is 12.81 $\mu m$ (NeII). For $\alpha_{s} \pm1$ this difference corresponds to 2 \% variation in flux densities. } 

Table \ref{tab:Photometry} lists both the reddened (F) and de-reddened (F') flux densities of all the reliable sources at both PAH1 ($8.59\mu m$) and NeII\_2 ($13.04 \mu m$), along with their spectral indices, which were calculated using, 
\begin{equation}
    \alpha_{s} = \frac{\log{(F'_{13.04}/F'_{8.59})}}{\log{(8.59/13.04)}}\,,
\end{equation}
i.e. using the convention $F\propto \nu^{+\alpha_{s}}$ or $F\propto \lambda^{-\alpha_{s}}$. In particular, hotter sources (non-embedded stars) are characterized by a positive spectral index in this convention, while dust-enshrouded stars, colder dusty filaments, or potential compact objects powered by non-thermal synchrotron emission (neutron stars) exhibit a steep power-law spectrum with a negative spectral index in the mid-infrared domain.

In Fig.~\ref{fig:Spectral_Index}, we depict the spectral indices of each source. The color and the size of each source indicate the spectral index and the flux density at the PAH1 band, respectively. In Fig.\ref{fig:Spectral_Index}, it is apparent that the brightest infrared-excess sources are also warm and blue, while the fainter and colder dust sources are along the mini-spiral.

\subsection{Cometary tail of IRS7}
\label{subsec:IRS7}

\begin{figure*}
\begin{center}
\includegraphics[width=0.75\textwidth, trim = 5 35 210 25, clip]{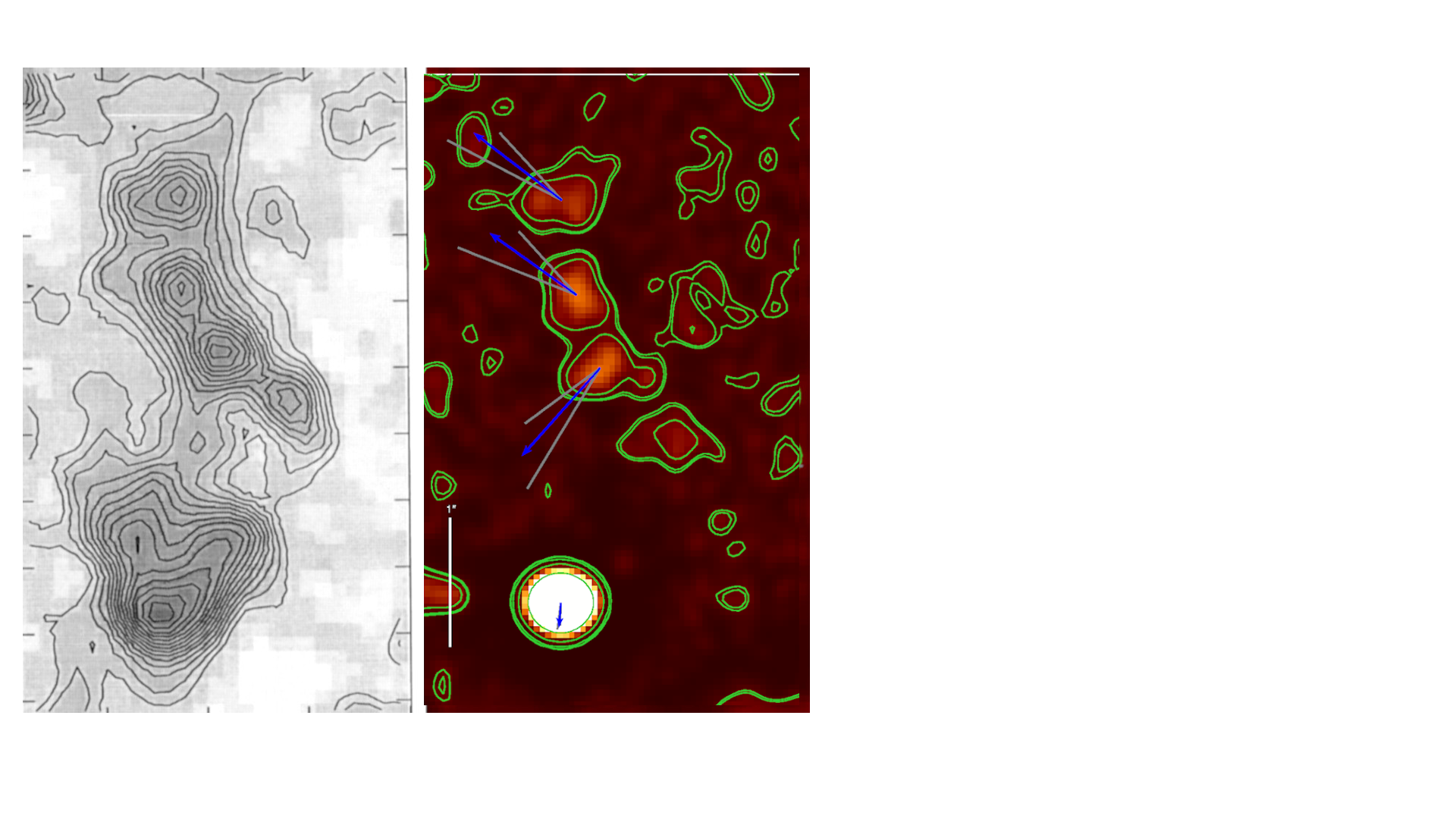}
\caption{On the left is an image of IRS7 and its extended tail at 2cm from \cite{Yusef-Zadeh1992ApJ...385L..41Y} and on the right is a smooth-subtracted image with the same FOV in PAH1 filter. The tail has 3 prominent substructures and their proper motions are indicated by blue arrows with grey lines as their error boundaries.}
\label{fig:IRS7Tail}
\end{center}
\end{figure*}

%
IRS7 (about 5.5" north of Sgr~A*) is one of the brightest infrared sources in the region and is classified as a pulsating M1/M2 red supergiant \citep{Carr2000ApJ...530..307C,Paumard2014A&A...568A..85P, Gravity2021A&A...651A..37G}. Radio \citep{Yusef-Zadeh1991ApJ...371L..59Y, Yusef-Zadeh1992ApJ...385L..41Y} and MIR \citep{Serabyn1991ApJ...378..557S} observations of IRS7  have revealed a bow shock feature towards the north and a cometary tail-like structure pointed directly away from Sgr~A*. Recently, \cite{Tsuboi2020PASJ...72...36T} also reported the shell-like  structure surrounding IRS7  and its northern extension in H30$\alpha$ recombination line. They measure line-of-sight velocities and conclude that the tail is the gas stream flowing from the shell around IRS7. As IRS7 moves southward, the pulsating release of gas as a stellar wind \citep{Paumard2014A&A...568A..85P} is left behind and is ionized by FUV radiation coming from the NSC. 

In the right panel of Fig.~\ref{fig:IRS7Tail}, we show the PAH1 (8.59$\mu$m) image of IRS7 and its extended tail in comparison with $\lambda$=2 cm contour map from \cite{Yusef-Zadeh1992ApJ...385L..41Y} which is on the left. Contours of individual substructures of the tail in both wavelengths seem to match well with each other, which is an indication of the quality of our data and filtering techniques. The tail extends from 1" to 3.6" north of IRS7 and has 3 major substructures. Blue arrows mark the proper motions and grey lines mark their error boundaries. The slight discrepancy in the direction of the transverse velocity in the MIR/NIR bands (see Appendix \ref{sec:ProperMotion_Comp}) and by using SiO maser astrometry \citep{Reid2003ApJ...587..208R,Borkar2020mbhe.confE..33B} could be caused by the fact that they probe different layers of the star. The proper motion measurements of the substructures of the tail reveal the dominant influences on each of them. The northern two which have detached from IRS7 earliest have almost similar velocities. This could mean that they are driven by the nuclear outflow coming from the inner few arcseconds.

In addition, the proper motion distribution of the {three} components reflects the combination of the downstream fluid motion in the IRS7 tail and the development of hydrodynamic instabilities, such as the Kelvin-Helmholtz (KH) instabilities, which are manifested as propagating waves moving transverse along the south-north direction of the flow, i.e. they introduce an additional turbulent velocity field to the predominant south-north downstream bow-shock flow. The development of the Kelvin-Helmholtz instability is expected due to the velocity shear between the shocked stellar wind of IRS7 and the surrounding hot medium. Assuming that the MIR-radio tail clumps formed due to the KH instability, we adopt their typical length-scale $\lambda_{\rm tail}\sim 0.5''\sim 0.02\,{\rm pc}$ from Fig.~\ref{fig:IRS7Tail}. The density and the temperature of the tail components, $n_{\rm tail}\sim 6\times 10^4\,{\rm cm^{-3}}$ and $T_{\rm tail}\sim 4650\,{\rm K}$, respectively, can be inferred from H30$\alpha$ recombination-line observations \citep{Tsuboi2020PASJ...72...36T}. These tail components are approximately in the pressure equilibrium with the surrounding hot medium, under the assumption of the extrapolation of the Bondi-radius values, $T_{\rm a}\sim 10^7\,{\rm K}$ and $n_{\rm a}\sim 26\,{\rm cm^{-3}}$ \citep{2003ApJ...591..891B}, i.e. $n_{\rm tail}T_{\rm tail}\sim n_{\rm a}T_{\rm a}$. This implies the density ratio between the ambient medium and the IRS7 tail of $r=n_{\rm a}/n_{\rm tail}\sim 4.3\times 10^{-4}$. For the shear velocity, we take the mean of the IRS7 stellar motions, $v_{\rm shear}\simeq v_{\star}\sim 180\,{\rm km\,s^{-1}}$, according to Table~\ref{tab:Comparision_pm}. The growth timescale of KH instabilities of the size $\lambda_{\rm tail}$ can be estimated as,   
\begin{equation}
    \tau_{\rm KH}\sim \frac{\lambda_{\rm tail}}{v_{\rm shear}}\frac{1+r}{\sqrt{r}}\sim 111\,{\rm yr}\,.
    \label{eq_KH_timescale}
\end{equation}
Since the crossing timescale of the IRS7 for the total length of the tail, $l_{\rm tail}\sim 3''$ is $\tau_{\rm cross}\sim l_{\rm tail}/v_{\star}\sim 652\,{\rm yr}\gtrsim \tau_{\rm KH}$, the KH instabilities of the length-scale of the observed clumps could have developed as IRS7 plows through the hot medium. The KH instabilities of the radius $R_{\rm tail}\sim 0.5\lambda_{\rm tail}$ can also survive long enough in the surrounding hot medium to be observed. For the pressure-confined colder clumps of the constant mass density $\rho_{\rm tail}$ that have the mass of $m_{\rm tail}=4/3\pi R_{\rm tail}^3 \rho_{\rm tail}$, the evaporation timescale is given by \citep{1977ApJ...211..135C,2021bhns.confE...1K},
\begin{align}
    \tau_{\rm evap}&=\frac{25k_{\rm B} n_{\rm tail}R_{\rm tail}^2}{8\kappa_{\rm H}}\,\notag\\
    &\simeq 4067\,\left(\frac{n_{\rm tail}}{6\times 10^4\,{\rm cm^{-3}}}\right)\left(\frac{R_{\rm tail}}{0.02\,{\rm pc}}\right)^2{\rm yr}  \,,
    \label{eq_evaporation_timescale}
\end{align}
where $\kappa_{\rm H}$ is the conductivity of the hot ambient medium, and its value can be calculated as $\kappa_{\rm H}=1.92\times 10^{11} {\rm erg\,s^{-1}\,K^{-1}\,cm^{-1}}$. Since $\tau_{\rm KH}<\tau_{\rm cross}<\tau_{\rm evap}$ the KH instabilities can properly explain the tail components of IRS7.

\subsection{Extended emission of IRS3}
\label{subsec:IRS3}

\begin{figure*}
\begin{center}
\includegraphics[width=0.99\textwidth, trim =0 0 20 0, clip]{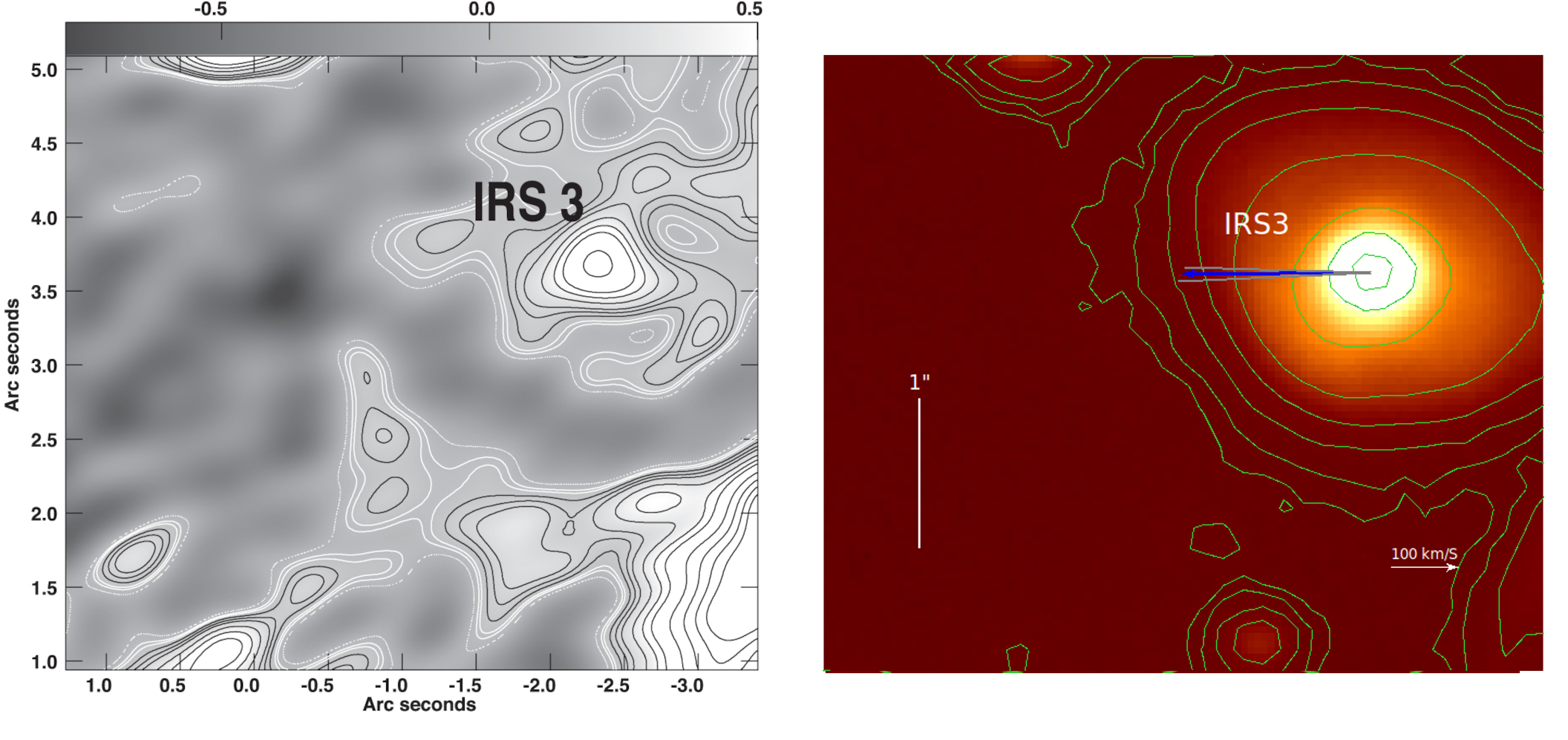}
\caption{On the left is a 226 GHz emission  of IRS3 and its surroundings from \cite{Yusef-Zadeh2017ApJ...837...93Y} and on the right is an untreated PAH1 (8.6 $\mu$m) image.  }
\label{fig:IRS3}
\end{center}
\end{figure*}

IRS3 is the brightest and the most extended stellar source at the Galactic center. Albeit the stellar core, the extended source is not circularly symmetric which has been a topic of interest. \cite{Viehmann2005A&A...433..117V} suggest that this morphology is due to a bow shock generated by winds from the massive central stellar cluster or an outflow from Sgr~A*. However, \cite{Yusef-Zadeh2017ApJ...837...93Y} suggests a tidal distortion scenario as IRS3 orbits around Sgr~A*. In Fig~\ref{fig:IRS3}, we compare the contour maps around the IRS3 with a 226 GHz image from \cite{Yusef-Zadeh2017ApJ...837...93Y}. 

By comparing the 226 GHz emission and the PAH1 image in Fig.~\ref{fig:IRS3}, the IRS3 circumstellar envelope exhibits similar structures at both wavelengths, in particular the extension in the north-east/south-west direction as well as another brighter extension to the south-east. Since the circumstellar material of IRS3 has an enormous radius of $R_{\rm IRS3}\sim 1''\sim 0.04\,{\rm pc}\sim 2\times 10^6\,R_{\odot}$, i.e. thousand times bigger than a red supergiant or an asymptotic giant branch star, it could be a signature of tidal elongation along the proper-motion direction as suggested by \citet{Yusef-Zadeh2017ApJ...837...93Y}. The question arises - what could have caused the expansion of the IRS3 envelope beyond its tidal (Hill) radius. If we consider the upper limit on the distance of IRS3 from Sgr~A* based on its mean stellar velocity of $v_{\star}\sim 266\,{\rm km\,s^{-1}}$, see Table~\ref{tab:Comparision_pm}, we obtain,
\begin{equation}
    d_{\rm IRS3}\lesssim 0.24\,{\rm pc}\,.
    \label{eq_distance_IRS3}
\end{equation}
Then we can estimate the upper limit on the tidal radius of IRS3,
\begin{align}
  R_{\rm tidal}&\lesssim d_{\rm IRS3}\left(\frac{m_{\rm IRS3}}{3M_{\bullet}} \right)^{1/3}\,\notag\\
  &\sim 10^5\left(\frac{d_{\rm IRS3}}{0.24\,{\rm pc}} \right) \left(\frac{m_{\rm IRS3}}{10\,M_{\odot}} \right)^{1/3}\left(\frac{M_{\bullet}}{4\times 10^6\,M_{\odot}} \right)^{-1/3}\,R_{\odot}\,.
\end{align}
Hence, even the largest red supergiants or AGB stars should have photospheres well inside their tidal radii at the distance of IRS3, see Eq.~\eqref{eq_distance_IRS3}. One possibility of the perturbation of the atmosphere of IRS3 is its ongoing or recent past interaction with the fast nuclear outflow or the jet, which led to the shock propagation inside the red supergiant envelope. This could effectively lead to the atmosphere ablation by the jet \citep{2020ApJ...903..140Z}. As a consequence, after one jet crossing, the IRS3 envelope adiabatically puffed up beyond the tidal (Hill) radius of the host star. The shocked envelope would expand adiabatically up to the tidal radius on the timescale given by the sound speed,
\begin{align}
    \tau_{\rm exp}&\sim \frac{R_{\rm tidal}}{c_{\rm s}}\,\notag\\
    &\simeq 172\,\left(\frac{R_{\rm tidal}}{10^5\,R_{\odot}} \right)\left(\frac{T_{\rm atm}}{10^4\,{\rm K}}\right)^{-1/2}\,{\rm yr}\,,
    \label{eq_expansion_timescale}
\end{align}
where $T_{\rm atm}$ is the temperature of the shocked red-giant envelope or the temperature of the red-giant photosphere after the jet ablation. 

Subsequently, after the short envelope expansion up to the tidal radius, the red-giant circumstellar material is further undergoing tidal prolongation along the orbital direction. The tidal stretching generally takes place on a timescale that is a fraction of the orbital timescale. The prolongation of the envelope of the radius $R_{\rm tidal}$ by a factor of $m=10$ is expected to occur in,
\begin{align}
    \tau_{\rm tidal}&=\sqrt{m}\frac{d_{\rm IRS3}^{3/2}}{\sqrt{GM_{\bullet}}}\,\\
    &=2772\,\left(\frac{m}{10}\right)^{1/2}\left(\frac{d_{\rm IRS3}}{0.24\,{\rm pc}}\right)^{3/2}\left(\frac{M_{\bullet}}{4\times 10^6\,M_{\bullet}}\right)^{-1/2}\text{years}\,.
    \label{eq_tidal_timescale}
\end{align}
Hence, the current extended state of IRS3 is likely the result of its atmosphere perturbation by the fast outflow or the jet, which led to the propagating shock inside the red giant envelope. This was followed by the fast envelope expansion up to the tidal radius, and subsequently the tidal prolongation along the IRS3 orbital motion took place. Both the processes are characterized by the timescales shorter than the orbital timescale, $P_{\rm IRS3}\sim 5508 (d_{\rm IRS3}/0.24\,{\rm pc})^{3/2}$ years, hence the current extended state of IRS3 is the result of a recent interaction a few thousand years ago.  

\subsection{Nature of dusty sources}
\label{subsec:NatureDustySources}

\begin{figure*}
\centering 
\subfloat{
  \includegraphics[width=0.45\textwidth,trim =25 6 45 40, clip]{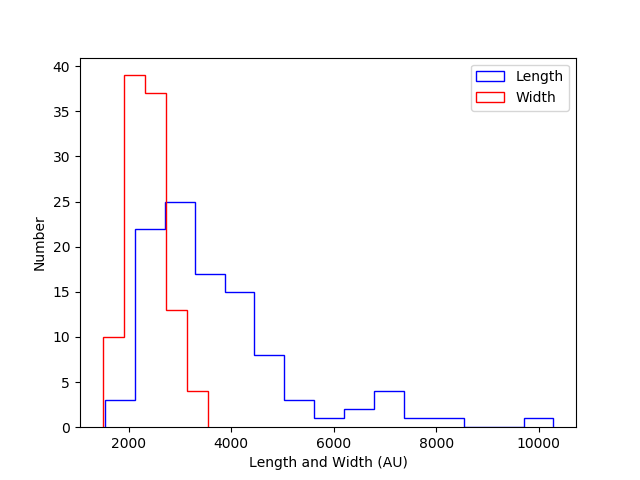}
}
\subfloat{
  \includegraphics[width=0.45\textwidth,trim =25 6 45 40, clip]{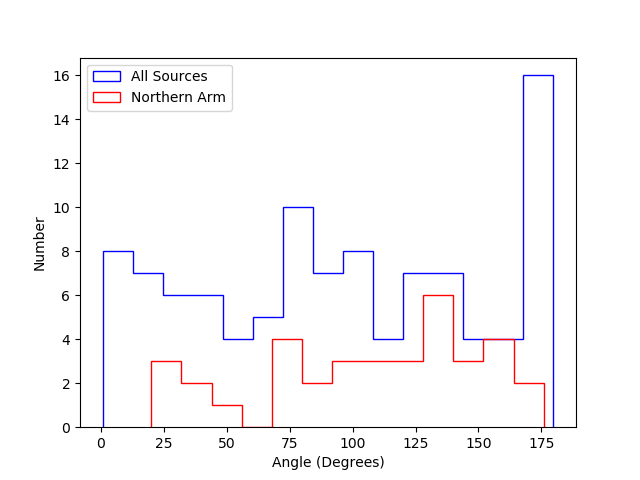}%
}
\quad
\subfloat{
  \includegraphics[width=0.45\textwidth,trim =25 6 45 40, clip]{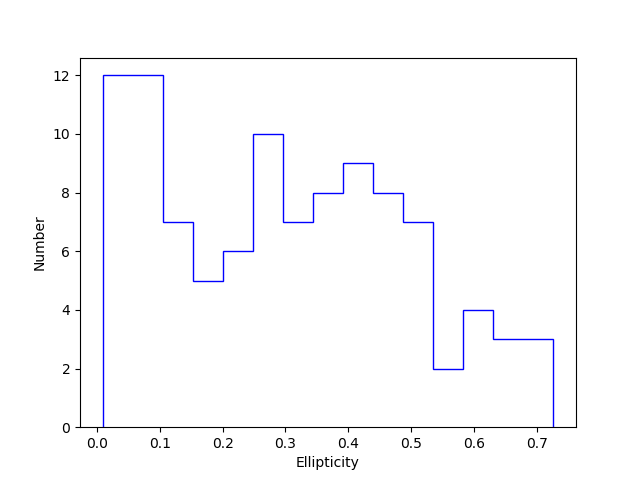}
}
\subfloat{
  \includegraphics[width=0.45\textwidth,trim =25 6 45 40, clip]{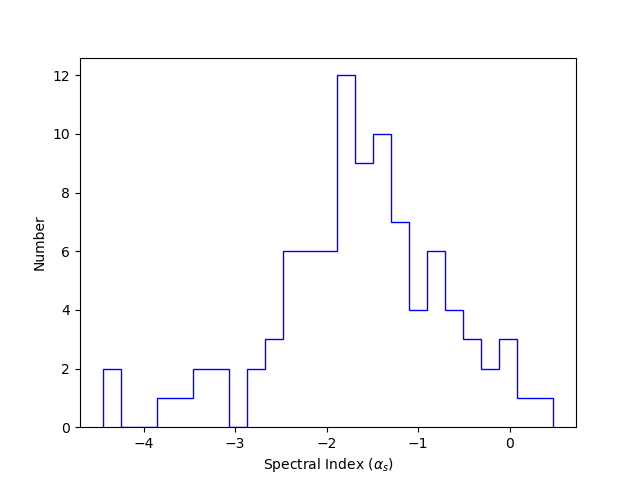}%
}
\caption{Histograms of physical properties of the extended dusty sources. Most sources are about $3500-4000$ AU long and  $2000-2500$ AU wide. Most sources are elliptic with various position angles. The sources on the northern arm seem to have a relatively tight spread of position angles compared to the rest of the sources. }
\label{fig:Histogram}
\end{figure*}

\begin{figure*}

\centering 
\subfloat{
  \includegraphics[width=0.4\textwidth,trim =4 6 45 40, clip]{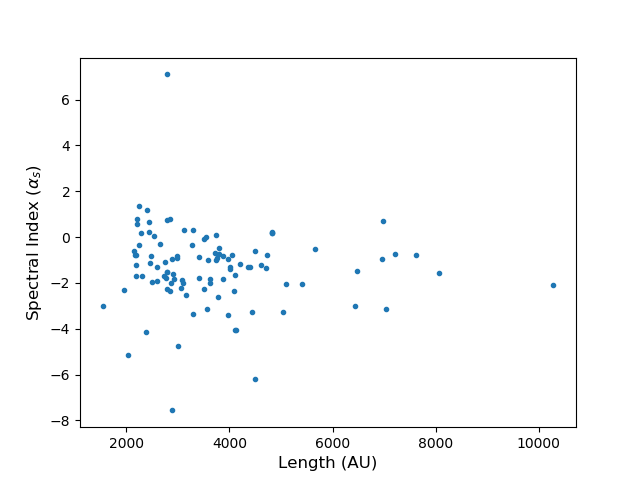}
}
\subfloat{
  \includegraphics[width=0.4\textwidth,trim =4 6 45 40, clip]{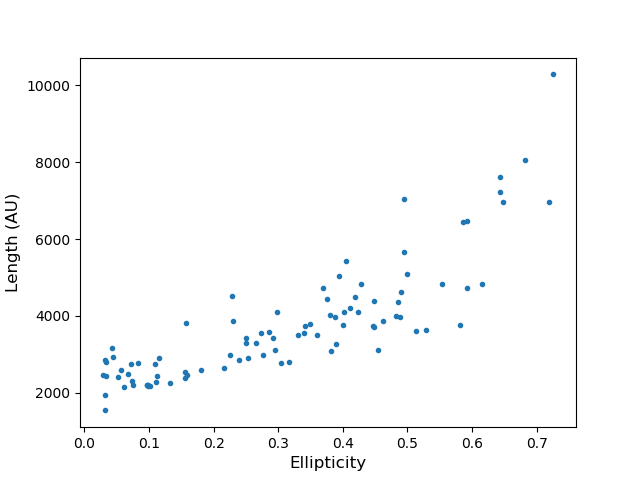}%
}
\quad
\subfloat{
  \includegraphics[width=0.4\textwidth,trim =4 6 45 40, clip]{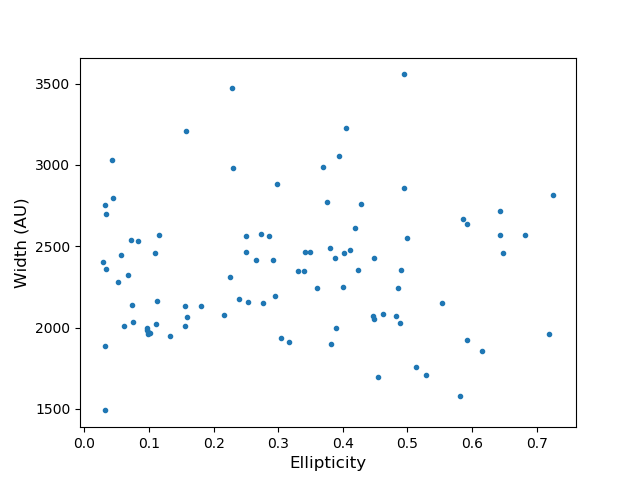}
}
\subfloat{
  \includegraphics[width=0.4\textwidth,trim = 4 6 45 40, clip]{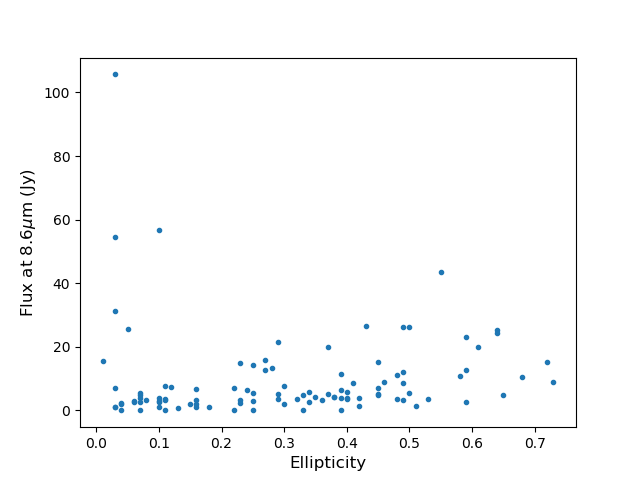}%
}
\quad
\subfloat{
  \includegraphics[width=0.4\textwidth,trim =4 6 45 40, clip]{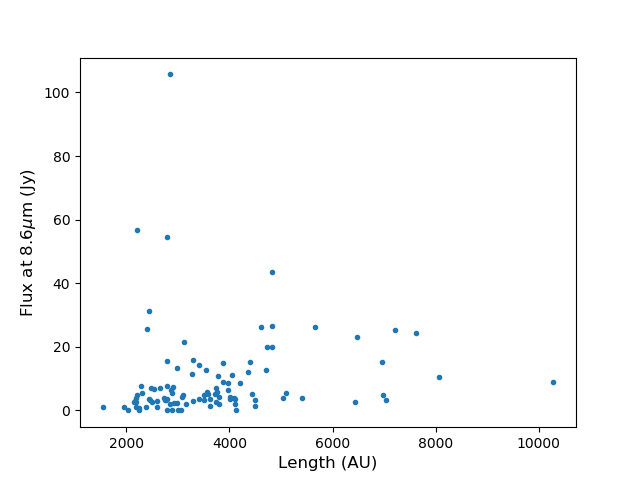}%
} 
 \subfloat{
  \includegraphics[width=0.4\textwidth,trim =4 6 45 40, clip]{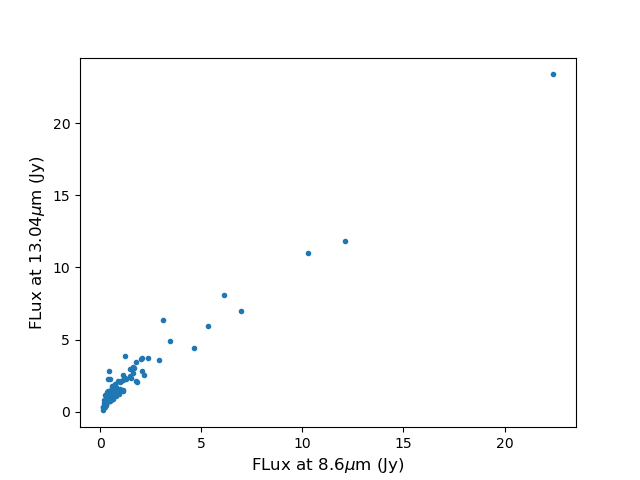}%
}

\quad
\subfloat{
  \includegraphics[width=0.4\textwidth,trim =4 6 45 40, clip]{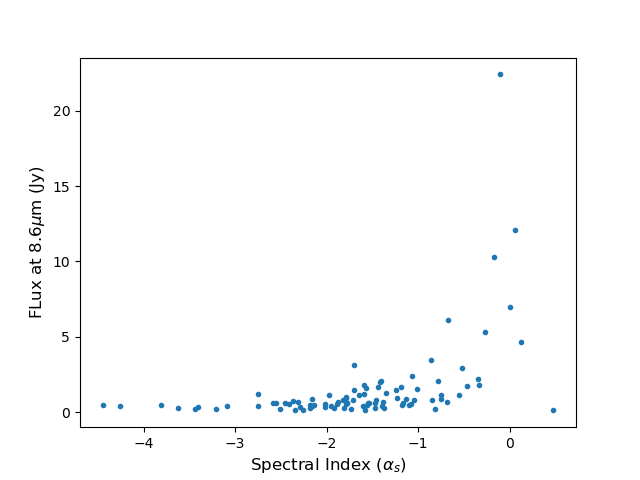}%
} 
 \subfloat{
  \includegraphics[width=0.4\textwidth,trim =4 6 45 40, clip]{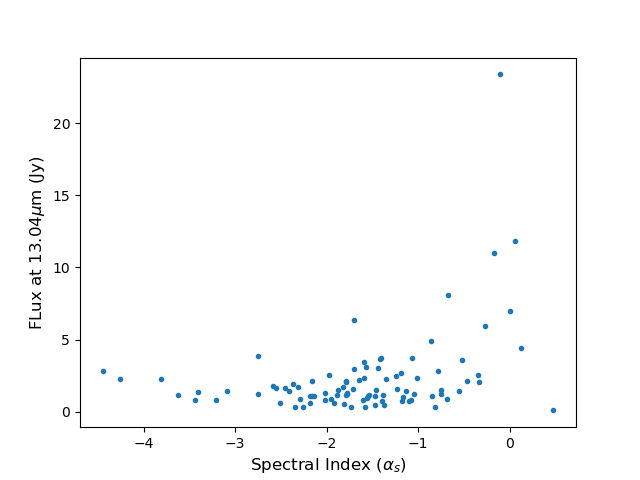}%
}
\caption{Correlation (or lack of it) among few physical properties of the extended dusty sources. The spectral index ($\alpha_{s}$) (using the convention $F\propto \nu^{+\alpha_{s}}$)  does not show a correlation with the length of the sources. Unsurprisingly, longer sources seem to be more elliptic. Flux densities at both the filters do correlate, however the spectral index has a tighter correlation with Flux at PAH1 than with Flux at NeII\_2. Although flux does not correlate with ellipticity or length, few noticeable outliers are those associated with bright IRS sources. }
\label{fig:Correlation}
\end{figure*}

\begin{table*}[]
\renewcommand\thetable{4}
    \centering
      \caption{Spearman correlation coefficients between quantities that characterize compact MIR dusty sources. In parantheses, we include the $p$-value. Five significant positive correlations were found: length-flux(NeII\_2), length-ellipticity, flux (PAH1)-flux(NeII\_2), flux (PAH1)-spectral index, and flux (NeII\_2)-ellipticity.}
    \resizebox{\textwidth}{!}{
    \begin{tabular}{c|c|c|c|c|c|c|c}
    \hline\hline
    & Length & Width & Angle & Flux (PAH1) & Flux (NeII\_2) & Spectral index & Ellipticity\\
    \hline
    Length        & -   & $0.064\, (0.524)$    & $-0.098\, (0.326)$  & $0.260\, (0.008)$  & $\mathbf{0.404\, (2.310\times 10^{-5})}$  & $-0.136\, (0.171)$  & $\mathbf{0.715\, (2.04 \times 10^{-17})}$  \\
    Width  & $0.064\, (0.524)$   & -    & $-0.091\,(0.361)$   & $0.146\,(0.140)$   & $0.235\,(0.017)$   & $-0.078\,(0.432)$  & $0.048\,(0.630)$   \\
    Angle  & $-0.098\, (0.326)$   & $-0.091\,(0.361)$    & -  & $-0.074\,(0.457)$  & $-0.180\,(0.069)$  & $0.089\, (0.372)$  & $-0.106\,(0.287)$    \\
    Flux (PAH1)   & $0.260\, (0.008)$     &  $0.146\,(0.140)$      & $-0.074\,(0.457)$   & -  & $\mathbf{0.850\,(7.36 \times 10^{-30})}$   & $\mathbf{0.510\,(3.65\times 10^{-8})}$  &  $0.307\,(0.0016)$  \\
    Flux (NeII\_2)   & $\mathbf{0.404\, (2.310\times 10^{-5})}$    & $0.235\,(0.017)$       &  $-0.180\,(0.069)$  & $\mathbf{0.850\,(7.36 \times 10^{-30})}$  & -  & $0.081\,(0.419)$   &  $\mathbf{0.382\,(6.70\times 10^{-5})}$  \\
    Spectral Index & $-0.136\, (0.171)$    & $-0.078\,(0.432)$   &  $0.089\, (0.372)$   & $\mathbf{0.510\,(3.65\times 10^{-8})}$    & $0.081\,(0.419)$    & -  &  $-0.073\,(0.466)$  \\
    Ellipticity   & $\mathbf{0.715\, (2.04 \times 10^{-17})}$    &  $0.048\,(0.630)$    & $-0.106\,(0.287)$   & $0.307\,(0.0016)$  &  $\mathbf{0.382\,(6.70\times 10^{-5})}$  & $-0.073\,(0.466)$    &  -  \\
    \hline
    \end{tabular}
    }
    \label{tab_correlation}
\end{table*}

To get a better understanding of the nature of the dusty sources in the GC, it becomes important to perform statistical analysis of all their physical properties. We fit a 2D-Gaussian on each source to determine its  length and width (major and minor axes of the elliptical aperture). In the top left image of Fig. \ref{fig:Histogram}, we show the histogram of these measurements. 

The distribution of position angles, measured anti-clockwise from the west is plotted in the top right image of Fig. \ref{fig:Histogram}. The sources along the northern arm have a smaller dispersion, which is expected given the fact they form a streamer-like structure. 

The ellipticity defined as $(a-b)/a$ where $a$ and $b$ are major and minor axes (length and width here), is plotted in the bottom left image of Fig. \ref{fig:Histogram}. Lastly, the bottom right image is a distribution of spectral indices. A large number of sources with negative indices reiterates the fact that the MIR region is dominated by dust-covered stars and colder dusty filaments. 

In Fig. \ref{fig:Correlation}, we try to find correlations between several fundamental physical properties of the dusty sources. In Table~\ref{tab_correlation}, we list the Spearman correlation coefficient $s$ as well as the corresponding $p$-value for all possible combinations of parameters, in total 21 pairs. We found { five significant positive correlations with sufficiently low $p$-values ($p<10^{-3}$)}: length - ellipticity, flux (NeII\_2) - length, flux (NeII\_2) - ellipticity, flux (PAH1) - flux (NeII\_2), and flux (PAH1) - spectral index. The positive correlation between the length and the ellipticity is expected due to {the definition of ellipticity}, while no significant correlation between the width and the ellipticity is caused by the narrow width distribution, which is caused by the lower cut-off due to the angular resolution, while the higher cut-off is likely determined rather by the physical process of the clump formation and evolution. The positive correlation between the flux densities at both wavelengths is also expected based on the spectral energy distribution for most of the dusty sources -- with the increasing wavelength, the flux density increases in the MIR domain. The positive correlation between the flux (PAH1) and the spectral index indicates that the differences among spectral energy distributions are mainly due to the flux density at $8.59\,{\rm \mu m}$ (PAH1) rather than due to the flux density at $13.04\,{\rm \mu m}$ (NeII\_2). The flatter spectrum is caused by the higher PAH1 flux density, while the NeII\_2-filter flux density is not significantly correlated with the spectral index. This is also visible in Fig.~\ref{fig:Spectral_Index}, where the most luminous sources in PAH1 filter have a clear tendency towards the positive spectral index (the bigger the symbol, the bluer the color). { This is also in line with the last two significant positive correlations: flux density (NeII\_2) and the length of the filaments, i.e. the longer the filament, the larger the flux density in the NeII\_2 filter. On the other hand, the correlation between the PAH1 flux density and the length is weaker and less significant. The positive correlation between the NeII\_2 flux density and the filament ellipticity then stems from the definition of the ellipticity that includes the length.}

The origin of the elongated blobs and filaments in the mini-spiral is uncertain. They could have arisen due to several processes that we list below:
\begin{itemize}
    \item[(a)] compression by the ram pressure produced by the collective stellar wind of the cluster of OB/Wolf-Rayet stars within the NSC. For a simple estimate, we assume the mass range of the clumps to be $m_{\rm clump}\approx 10^{-2}-10^{-4}\,M_{\odot}$. We further assume the ambient number density of $n_{\rm a}\sim 10 {\rm cm^{-3}}$, i.e. close to the Bondi-sphere value, and the outflow stellar wind velocity of $v_{\rm w}\sim 1000\,{\rm km\,s^{-1}}$ \citep{1997A&A...325..700N}. For the stationary set-up, in which the MIR filaments are pressure-confined by the wind pressure acting on the mini-spiral clumps moving with the orbital velocity of $v_{\rm clump}\sim 100\,{\rm km\,s^{-1}}$ \citep{2009ApJ...699..186Z}, we may obtain their characteristic length-scale using,
    \begin{align}
        \lambda_{\rm clump}\sim \left(\frac{m_{\rm clump}k_{\rm B}T_{\rm clump}}{\mu m_{\rm H}[\mu m_{\rm H}(v_{\rm clump}^2+v_{\rm w}^2)-B^2/8\pi] }\right)^{1/3}\,,
    \end{align}
    where the ram pressure is balanced by the thermal pressure as well as the magnetic field inside the filaments. For the clump temperature of $T_{\rm clump}\sim 10^4 {\rm K}$ \citep{2010ApJ...723.1097Z} and a negligible magnetic field, the characteristic length-scale is $\lambda_{\rm clump}\sim 10\,500 - 2\,300\,{\rm AU}$, which captures the range for the inferred length and the width of the MIR filaments, see Fig.~\ref{fig:Histogram} (top left panel). An aligned magnetic field in the Northern arm, see e.g. \citet{1998MNRAS.299..743A,2018MNRAS.476..235R}, has a tendency to increase the length-scale of the clumps; 

    \item[(b)] the blobs are the manifestation of the Kelvin-Helmholtz instability that arises due to the streaming motion of the mini-spiral in the surrounding hot plasma. This is especially the case for the Northern and the Eastern arms that have more radial, elliptical orbits in comparison with the Western arm \citep{2009ApJ...699..186Z}. For the estimate of the characteristic growth timescale of the KH instabilities of a certain size $\lambda_{\rm clump}\sim 1000-10\,000\,{\rm AU}\sim 1.5\times 10^{16}-1.5\times 10^{17}\,{\rm cm}$, we will assume the shearing velocity given by the terminal stellar wind velocity that is approximately perpendicular to the mini-spiral orbital motion, i.e. $v_{\rm w}\sim 1000\,{\rm km\,s^{-1}}$ and $v_{\rm mini}\sim 100\,{\rm km\,s^{-1}}$, which gives $v_{\rm shear}\sim \sqrt{v_{\rm w}^2+v_{\rm mini}^2}\sim 1000\,{\rm km\,s^{-1}}$. The density of the ionized component of the mini-spiral was inferred to be $n_{\rm mini}\sim 3-21 \times 10^4\,{\rm cm^{-3}}$ \citep{2010ApJ...723.1097Z}, which sets the ratio with respect to the ambient hot plasma to $r=n_{\rm a}/n_{\rm mini}\sim 4.8\times 10^{-5}-3.3\times 10^{-4}$ assuming $n_{\rm a}\sim 10\,{\rm cm^{-3}}$. Using Eq.~\eqref{eq_KH_timescale}, we get the KH growth timescale of $\tau_{\rm KH}(1000\,{\rm AU})\sim 262- 686$ years and $\tau_{\rm KH}(10000\,{\rm AU})\sim 2620- 6860$ years. Since the lifetime of the mini-spiral is $\sim 6\times 10^4$ years, as given by the mean orbital timescale of the arms \citep{2009ApJ...699..186Z}, the clumps of $\lambda_{\rm clump}\sim 1000\,{\rm AU}$ are continually forming as well as disappearing along the arms since their lifetime is comparable to the formation time as given by the evaporation timescale, see Eq.~\eqref{eq_evaporation_timescale}, which gives $\tau_{\rm evap}\sim 480-3360$ years for the clump size of $\lambda_{\rm clump}\sim 1000\,{\rm AU}$, the density range of $n_{\rm clump}\sim 3-21\times 10^4\,{\rm cm^{-3}}$, and the conductivity of the hot medium $\kappa_{\rm H}\sim 1.9\times 10^{11} T_{7}^{5/2}\,{\rm erg\,s^{-1}\,K^{-1}\,cm^{-1}}$ where $T_{7}$ is in units of $10^7\,{\rm K}$. This is consistent with the early considerations of the transient nature of ionized clouds in Sgr~A West, see \citet{1980ApJ...241..132L};
  
     \item[(c)] the denser clumps could have formed as a result of the thermal instability during the past states of a higher activity of Sgr~A* when its bolometric luminosity reached $\sim 10^{39}-10^{41}\,{\rm erg\,s^{-1}}$. This was proposed by \citet{2014MNRAS.445.4385R}, who studied the conditions for the thermal instability in the mini-spiral region. They concluded that the thermal instability does not operate in the current low-luminous state of Sgr~A*, but during the past periods of enhanced activity, the thermal instability likely operated up to $\sim 1.4\,{\rm pc}$ from Sgr~A*, which led to the formation of a two-phase medium. {This increased the clumpiness of the mini-spiral by creating clumps of $\sim 1-100$ Earth masses, i.e. $3\times 10^{-6}-3\times 10^{-4}\,M_{\odot}$. Based on the minispiral clump mass estimate presented in Section~\ref{sec:Discussion}, the thermal instability clump mass appears to be lower by at least two orders magnitude.} {In case the thermal instability operated a few hundred years ago and clumps with the radius of $R_{\rm clump}\sim 10^{14}-10^{15}\,{\rm cm}$ and the mass of $m_{\rm clump}\sim 3\times 10^{-5}\,M_{\odot}$ formed, then the evaporation timescale according to Eq.~\eqref{eq_evaporation_timescale} is long enough ($\sim 10^3-10^4$ years) so that they could in principle survive till nowadays}. {However}, by including the mechanical heating of the NSC, \citet{2017MNRAS.464.2090R} concluded that cold clumps are not expected to form in the vicinity of Sgr~A* since stellar winds induce the outflow of the hot ambient plasma. {This fact, in combination with the small clump mass, renders the instability scenario less plausible to explain the formation of the elongated MIR sources, unless each MIR filament would consist of at least 100 clumps formed via the thermal instability. However, since the models of \citet{2014MNRAS.445.4385R} and \citet{2017MNRAS.464.2090R} assume a spherical symmetry it is difficult to explain the filamentary structure of the instability clumps in this framework}.   
\end{itemize}
Regardless of the way how the denser and the brighter MIR clumps formed, {their shape and orientation} are the result of the tidal prolongation along the streaming motion of the Northern arm. This stems from the tidal radius estimated for the mini-spiral clump of radius $R_{\rm clump}\gtrsim 1000\,{\rm AU}$ and mass $m_{\rm clump}\leq 0.1\,M_{\odot}$,
\begin{align}
    r_{\rm t}&\simeq R_{\rm clump}\left(\frac{2M_{\bullet}}{m_{\rm clump}}\right)^{1/3}\,\notag\\
    &\sim 2.1\,\left(\frac{R_{\rm clump}}{1000\,{\rm AU}}\right)\left(\frac{m_{\rm clump}}{0.1\,M_{\odot}}\right)^{-1/3}\left(\frac{M_{\bullet}}{4\times 10^6\,M_{\odot}}\right)^{1/3}{\rm pc}\,.
    \label{eq_tidal_radius_clump}
\end{align}
Since the mini-spiral arms are closer than $2\,{\rm pc}$, any clump inside them is susceptible to tidal forces and prolongation parallel to the streaming (orbital) motion of the arms.


\section{Discussion}
\label{sec:Discussion}

Here we present, to our knowledge, the largest consistent number of individual source identifications, proper motions and thus tangential velocities, and MIR spectral indices in the central parsec of the Galaxy, covering the central stellar cluster and the immediate Sgr~A* supermassive black hole region. The MIR sources can clearly be divided into two groups: one associated with the flow along the extended mini-spiral and one associated with dust-enshrouded stars and infrared-excess sources of the central stellar cluster.

Based on the color-temperature map derived from 12.5~$\mu$m and 20.3~$\mu$m maps by \citet{Cotera1999ASPC..186..240C}, the overall temperature in the mini-spiral is of the order of $\sim$200 K. We assume that this is a good estimate or at least a lower limit to the temperatures of the infrared-excess sources within the mini-spiral. A more exact estimate of the temperatures has to await the multi-frequency mid-infrared data at a high angular resolution, as will be obtained e.g. with the METIS instrument at the ELT \citep{Brandl2021Msngr.182...22B}.

For the sources not associated with stars, we estimate their dust masses based on PAH1 and NeII\_2 flux densities. We use the relation based on \citet{1978ApJ...220..556R} and \citet{Kunneriath2012A&A...538A.127K},
\begin{equation}
   M_{\rm d}=\frac{F(\nu)D_{\rm GC}^2}{B(\nu, T_{\rm d})}\frac{4a}{3Q(\nu)}\rho_{\rm d}\,,
   \label{eq_dust_mass}
\end{equation}
where $F(\nu)$ is the measured flux density and $B(\nu, T_{\rm d})$ is the Planck function calculated for the dust temperature of $T_{\rm d}\sim 200\,{\rm K}$ \citep{Cotera1999ASPC..186..240C}. The dust is characterized by the mean values of the radius $a=0.1\,{\rm \mu m}$, the mass density $\rho_{\rm d}=2\,{\rm g\,cm^{-3}}$, and the emissivity $Q\approx 10^{-3}-10^{-2}$ close to $10\,{\rm \mu m}$ \citep{1975ApJ...200...30A,1978ApJ...220..556R}. The distance to the Galactic center is set to $D_{\rm GC}=8.1\,{\rm kpc}$. Assuming the gas-to-dust ratio of $\sim 100$, {{we obtain the mean gas mass of a mini-spiral clump, $\overline{M}_{\rm g}({\rm PAH1})\sim 0.046-0.46\,M_{\odot}$ and  $\overline{M}_{\rm g}({\rm NeII\_2})\sim 0.015-0.15\,M_{\odot}$ as based on 85 dusty filaments that are not associated with stellar sources. These values are consistent within a factor of three. Considering the peak width of the clumps, $w\sim 2000\,{\rm AU}$, and the peak length, $l\sim 3000\,{\rm AU}$, we obtain the characteristic clump volume of $V_{\rm clump}\sim 3.2\times 10^{49}\,{\rm cm^3}$, which yields the filament number density in the range of $n_{\rm clump}\sim 1.1\times 10^6-3.4\times 10^7\,{\rm cm^{-3}}$, which is at least one order of magnitude more than the electron number density of the ionized component $n_{\rm e}=(3-21)\times 10^{4}\,{\rm cm^{-3}}$ \citep{2010ApJ...723.1097Z}. The filaments could thus be overdensity regions that are either pressure-confined by the stellar winds of OB/WR stars or they could stand for KH instabilities that got denser due to radiative cooling. This supports the multiphase nature of the minispiral streamers, with denser filaments embedded within a more diluted ionized gas \citep{2014MNRAS.445.4385R}. Denser dusty filaments could also be the sites of the water- and CO-ice features and hydrocarbons detected within the central parsec \citep{2015ApJ...806..202M,2015MNRAS.448.3363M}. If the mean clump gas mass is in the range of $\sim 0.01-0.1\,M_{\odot}$ and we have $\sim 100$ filaments, then their total gas mass of $\sim 1-10\,M_{\odot}$ is consistent with the total ionized gas mass of $\sim 60\,M_{\odot}$ within the central cavity \citep{Lo1983Natur.306..647L}. The denser filaments are currently not massive and dense enough to form stellar and sub-stellar objects. As discussed in the previous section, they are transient features formed via the KH instability along the streaming motion and on the timescale of $\sim 100-1000$ years, they evaporate. The filaments are also expected to be tidally elongated along the streaming motion of the minispiral during their lifetime.}}

\cite{Muzic2007A&A...469..993M} show that the shape and the motion of the mini-spiral filaments do not agree with a purely Keplerian motion of the gas in the potential of the supermassive black hole at the position of Sgr~A*. The authors involve additional mechanisms that are responsible for the formation and the motion of these filaments. They assume that the filaments are affected by an outflow from the disk of young mass-losing stars around Sgr~A*. In addition, an outflow from the Sgr~A* black hole region itself may be responsible for the elongated shape and the motion of the filaments.

\section{Summary}
\label{sec:Summary}

We studied MIR images of the central parsec of the Galactic center in the $N$-band (8.6 $\mu$m and 13.04 $\mu$m). As the MIR emission is dominated by dust and extended regions around the central SMBH, we applied a high-pass filter on the images to resolve and identify the sources. We present the proper motions of these extended objects over 12 year-time period. There are two distinct types of the observed motion; one related to infrared-excess sources of the central stellar cluster and the other, stream-like motion of extended objects along the mini-spiral streamer. We also present the flux densities of all the sources using elliptical apertures. Using the spectral indices, we infer that the MIR region is dominated by the dust enshrouded stars or colder dusty filaments and the temperature of $\sim$ 200 K \citep{Cotera1999ASPC..186..240C} is at least the lower limit of infrared-excess sources within the mini-sipiral. We detect the bow-shock feature and the tail components of IRS7 that are pointed away from Sgr~A*. Proper motion distribution of the individual tail components can be interpreted with the combination of downstream fluid motion and the development of KH instabilities. We detect and resolve the brightest MIR source in the region IRS3. The extended structure of the star is likely a result of its atmosphere perturbation followed by tidal prolongation. We also report on the nature of all the dusty sources and delve into their possible origins.

\section*{Acknowledgements}
This work was supported in part by SFB 956—Conditions and Impact of Star Formation. HB and EH are members of the International Max Planck Research School for Astronomy and Astrophysics at the Universities of Bonn and Cologne.
We thank the Collaborative Research Centre 956, sub-project A02, funded by the Deutsche Forschungsgemeinschaft (DFG) – project ID 184018867. NBS acknowledges financial support from the Austrian National Science Foundation through FWF stand-alone grant P31154-N27. MZ acknowledges the financial support by the GA\v{C}R EXPRO grant No. 21-13491X ``Exploring the Hot Universe and Understanding Cosmic Feedback". 
RS acknowledges financial support from the State Agency for Research of the Spanish MCIU through the “Center of Excellence Severo Ochoa” award for the Instituto de Astrof\'{i}sica de Andaluc\'{i}a (SEV-2017- 0709) and financial support from national project PGC2018-095049-B-C21 (MCIU/AEI/FEDER, UE).


\bibliography{sample631}{}

\begin{thebibliography}{}
\expandafter\ifx\csname natexlab\endcsname\relax\def\natexlab#1{#1}\fi
\providecommand{\url}[1]{\href{#1}{#1}}
\providecommand{\dodoi}[1]{doi:~\href{http://doi.org/#1}{\nolinkurl{#1}}}
\providecommand{\doeprint}[1]{\href{http://ascl.net/#1}{\nolinkurl{http://ascl.net/#1}}}
\providecommand{\doarXiv}[1]{\href{https://arxiv.org/abs/#1}{\nolinkurl{https://arxiv.org/abs/#1}}}

\bibitem[{{Aannestad}(1975)}]{1975ApJ...200...30A}
{Aannestad}, P.~A. 1975, \apj, 200, 30, \dodoi{10.1086/153757}

\bibitem[{{Aitken} {et~al.}(1998){Aitken}, {Smith}, {Moore}, \&
  {Roche}}]{1998MNRAS.299..743A}
{Aitken}, D.~K., {Smith}, C.~H., {Moore}, T. J.~T., \& {Roche}, P.~F. 1998,
  \mnras, 299, 743, \dodoi{10.1046/j.1365-8711.1998.01807.x}

\bibitem[{{Baganoff} {et~al.}(2003){Baganoff}, {Maeda}, {Morris}, {Bautz},
  {Brandt}, {Cui}, {Doty}, {Feigelson}, {Garmire}, {Pravdo}, {Ricker}, \&
  {Townsley}}]{2003ApJ...591..891B}
{Baganoff}, F.~K., {Maeda}, Y., {Morris}, M., {et~al.} 2003, \apj, 591, 891,
  \dodoi{10.1086/375145}

\bibitem[{{Becklin} {et~al.}(1982){Becklin}, {Gatley}, \&
  {Werner}}]{Becklin1982ApJ...258..135B}
{Becklin}, E.~E., {Gatley}, I., \& {Werner}, M.~W. 1982, \apj, 258, 135,
  \dodoi{10.1086/160060}

\bibitem[{{Blandford} \& {Begelman}(1999)}]{Blandford1999MNRAS.303L...1B}
{Blandford}, R.~D., \& {Begelman}, M.~C. 1999, \mnras, 303, L1,
  \dodoi{10.1046/j.1365-8711.1999.02358.x}

\bibitem[{{Blank} {et~al.}(2016){Blank}, {Morris}, {Frank},
  {Carroll-Nellenback}, \& {Duschl}}]{2016MNRAS.459.1721B}
{Blank}, M., {Morris}, M.~R., {Frank}, A., {Carroll-Nellenback}, J.~J., \&
  {Duschl}, W.~J. 2016, \mnras, 459, 1721, \dodoi{10.1093/mnras/stw771}

\bibitem[{{Blum} {et~al.}(1996){Blum}, {Sellgren}, \&
  {Depoy}}]{Blum1996ApJ...470..864B}
{Blum}, R.~D., {Sellgren}, K., \& {Depoy}, D.~L. 1996, \apj, 470, 864,
  \dodoi{10.1086/177917}

\bibitem[{{Borkar} {et~al.}(2020){Borkar}, {Eckart}, {Straubmeier}, {Sabha},
  {Sjouwerman}, {Karas}, {Kunneriath}, {Moser}, {Britzen},
  {Valencia-Schneider}, {Donea}, \& {Zensus}}]{Borkar2020mbhe.confE..33B}
{Borkar}, A., {Eckart}, A., {Straubmeier}, C., {et~al.} 2020, in Multifrequency
  Behaviour of High Energy Cosmic Sources - XIII. 3-8 June 2019. Palermo, 33.
\newblock \doarXiv{1909.13753}

\bibitem[{{Bower} {et~al.}(2003){Bower}, {Wright}, {Falcke}, \&
  {Backer}}]{Bower2003ApJ...588..331B}
{Bower}, G.~C., {Wright}, M. C.~H., {Falcke}, H., \& {Backer}, D.~C. 2003,
  \apj, 588, 331, \dodoi{10.1086/373989}

\bibitem[{{Brandl} {et~al.}(2021){Brandl}, {Bettonvil}, {van Boekel},
  {Glauser}, {Quanz}, {Absil}, {Amorim}, {Feldt}, {Glasse}, {G{\"u}del}, {Ho},
  {Labadie}, {Meyer}, {Pantin}, {van Winckel}, \& {METIS
  Consortium}}]{Brandl2021Msngr.182...22B}
{Brandl}, B., {Bettonvil}, F., {van Boekel}, R., {et~al.} 2021, The Messenger,
  182, 22, \dodoi{10.18727/0722-6691/5218}

\bibitem[{{Carr} {et~al.}(2000){Carr}, {Sellgren}, \&
  {Balachandran}}]{Carr2000ApJ...530..307C}
{Carr}, J.~S., {Sellgren}, K., \& {Balachandran}, S.~C. 2000, \apj, 530, 307,
  \dodoi{10.1086/308340}

\bibitem[{{Christopher} {et~al.}(2005){Christopher}, {Scoville}, {Stolovy}, \&
  {Yun}}]{Christopher2005ApJ...622..346C}
{Christopher}, M.~H., {Scoville}, N.~Z., {Stolovy}, S.~R., \& {Yun}, M.~S.
  2005, \apj, 622, 346, \dodoi{10.1086/427911}

\bibitem[{{Cotera} {et~al.}(1999){Cotera}, {Morris}, {Ghez}, {Becklin},
  {Tanner}, {Werner}, \& {Stolovy}}]{Cotera1999ASPC..186..240C}
{Cotera}, A., {Morris}, M., {Ghez}, A.~M., {et~al.} 1999, in Astronomical
  Society of the Pacific Conference Series, Vol. 186, The Central Parsecs of
  the Galaxy, ed. H.~{Falcke}, A.~{Cotera}, W.~J. {Duschl}, F.~{Melia}, \&
  M.~J. {Rieke}, 240

\bibitem[{{Cowie} \& {McKee}(1977)}]{1977ApJ...211..135C}
{Cowie}, L.~L., \& {McKee}, C.~F. 1977, \apj, 211, 135, \dodoi{10.1086/154911}

\bibitem[{{Eckart} {et~al.}(1992){Eckart}, {Genzel}, {Krabbe}, {Hofmann}, {van
  der Werf}, \& {Drapatz}}]{Eckart1992Natur.355..526E}
{Eckart}, A., {Genzel}, R., {Krabbe}, A., {et~al.} 1992, \nat, 355, 526,
  \dodoi{10.1038/355526a0}

\bibitem[{{Eckart} {et~al.}(2002){Eckart}, {Genzel}, {Ott}, \&
  {Sch{\"o}del}}]{Ecakrt2002MNRAS.331..917E}
{Eckart}, A., {Genzel}, R., {Ott}, T., \& {Sch{\"o}del}, R. 2002, \mnras, 331,
  917, \dodoi{10.1046/j.1365-8711.2002.05237.x}

\bibitem[{{Eckart} {et~al.}(2017){Eckart}, {H{\"u}ttemann}, {Kiefer},
  {Britzen}, {Zaja{\v{c}}ek}, {L{\"a}mmerzahl}, {St{\"o}ckler}, {Valencia-S},
  {Karas}, \& {Garc{\'\i}a-Mar{\'\i}n}}]{2017FoPh...47..553E}
{Eckart}, A., {H{\"u}ttemann}, A., {Kiefer}, C., {et~al.} 2017, Foundations of
  Physics, 47, 553, \dodoi{10.1007/s10701-017-0079-2}

\bibitem[{{Falcke} \& {Markoff}(2013)}]{2013CQGra..30x4003F}
{Falcke}, H., \& {Markoff}, S.~B. 2013, Classical and Quantum Gravity, 30,
  244003, \dodoi{10.1088/0264-9381/30/24/244003}

\bibitem[{{Fritz} {et~al.}(2011){Fritz}, {Gillessen}, {Dodds-Eden}, {Lutz},
  {Genzel}, {Raab}, {Ott}, {Pfuhl}, {Eisenhauer}, \&
  {Yusef-Zadeh}}]{Fritz2011ApJ...737...73F}
{Fritz}, T.~K., {Gillessen}, S., {Dodds-Eden}, K., {et~al.} 2011, \apj, 737,
  73, \dodoi{10.1088/0004-637X/737/2/73}

\bibitem[{{Genzel} {et~al.}(2010){Genzel}, {Eisenhauer}, \&
  {Gillessen}}]{2010RvMP...82.3121G}
{Genzel}, R., {Eisenhauer}, F., \& {Gillessen}, S. 2010, Reviews of Modern
  Physics, 82, 3121, \dodoi{10.1103/RevModPhys.82.3121}

\bibitem[{{Genzel} {et~al.}(2000){Genzel}, {Pichon}, {Eckart}, {Gerhard}, \&
  {Ott}}]{Genzel2000MNRAS.317..348G}
{Genzel}, R., {Pichon}, C., {Eckart}, A., {Gerhard}, O.~E., \& {Ott}, T. 2000,
  \mnras, 317, 348, \dodoi{10.1046/j.1365-8711.2000.03582.x}

\bibitem[{{Gravity Collaboration} {et~al.}(2021){Gravity Collaboration},
  {Rodr{\'\i}guez-Coira}, {Paumard}, {Perrin}, {Vincent}, {Abuter}, {Amorim},
  {Baub{\"o}ck}, {Berger}, {Bonnet}, {Brandner}, {Cl{\'e}net}, {de Zeeuw},
  {Dexter}, {Drescher}, {Eckart}, {Eisenhauer}, {F{\"o}rster Schreiber}, {Gao},
  {Garcia}, {Gendron}, {Genzel}, {Gillessen}, {Habibi}, {Haubois}, {Henning},
  {Hippler}, {Horrobin}, {Jimenez-Rosales}, {Jochum}, {Jocou}, {Kaufer},
  {Kervella}, {Lacour}, {Lapeyr{\`e}re}, {Le Bouquin}, {L{\'e}na}, {Nowak},
  {Ott}, {Perraut}, {Pfuhl}, {Sanchez-Bermudez}, {Shangguan}, {Scheithauer},
  {Stadler}, {Straub}, {Straubmeier}, {Sturm}, {Tacconi}, {Shimizu}, {von
  Fellenberg}, {Waisberg}, {Widmann}, {Wieprecht}, {Wiezorrek}, {Woillez},
  {Yazici}, \& {Zins}}]{Gravity2021A&A...651A..37G}
{Gravity Collaboration}, {Rodr{\'\i}guez-Coira}, G., {Paumard}, T., {et~al.}
  2021, \aap, 651, A37, \dodoi{10.1051/0004-6361/202039501}

\bibitem[{{Hsieh} {et~al.}(2021){Hsieh}, {Koch}, {Kim}, {Mart{\'\i}n}, {Yen},
  {Carpenter}, {Harada}, {Turner}, {Ho}, {Tang}, \&
  {Beck}}]{Hsieh2021ApJ...913...94H}
{Hsieh}, P.-Y., {Koch}, P.~M., {Kim}, W.-T., {et~al.} 2021, \apj, 913, 94,
  \dodoi{10.3847/1538-4357/abf4cd}

\bibitem[{{Jackson} {et~al.}(1993){Jackson}, {Geis}, {Genzel}, {Harris},
  {Madden}, {Poglitsch}, {Stacey}, \& {Townes}}]{Jackson1993ApJ...402..173J}
{Jackson}, J.~M., {Geis}, N., {Genzel}, R., {et~al.} 1993, \apj, 402, 173,
  \dodoi{10.1086/172120}

\bibitem[{{Karas} {et~al.}(2021){Karas}, {Svoboda}, \&
  {Zaja{\v{c}}ek}}]{2021bhns.confE...1K}
{Karas}, V., {Svoboda}, J., \& {Zaja{\v{c}}ek}, M. 2021, in RAGtime: Workshops
  on black holes and netron stars, E1.
\newblock \doarXiv{1901.06507}

\bibitem[{{Kunneriath} {et~al.}(2012){Kunneriath}, {Eckart}, {Vogel}, {Teuben},
  {Mu{\v{z}}i{\'c}}, {Sch{\"o}del}, {Garc{\'\i}a-Mar{\'\i}n}, {Moultaka},
  {Staguhn}, {Straubmeier}, {Zensus}, {Valencia-S.}, \&
  {Karas}}]{Kunneriath2012A&A...538A.127K}
{Kunneriath}, D., {Eckart}, A., {Vogel}, S.~N., {et~al.} 2012, \aap, 538, A127,
  \dodoi{10.1051/0004-6361/201117676}

\bibitem[{{Lacy} {et~al.}(1980){Lacy}, {Townes}, {Geballe}, \&
  {Hollenbach}}]{1980ApJ...241..132L}
{Lacy}, J.~H., {Townes}, C.~H., {Geballe}, T.~R., \& {Hollenbach}, D.~J. 1980,
  \apj, 241, 132, \dodoi{10.1086/158324}

\bibitem[{{Lo} \& {Claussen}(1983)}]{Lo1983Natur.306..647L}
{Lo}, K.~Y., \& {Claussen}, M.~J. 1983, \nat, 306, 647,
  \dodoi{10.1038/306647a0}

\bibitem[{{Lutz} {et~al.}(1993){Lutz}, {Krabbe}, \&
  {Genzel}}]{Lutz1993ApJ...418..244L}
{Lutz}, D., {Krabbe}, A., \& {Genzel}, R. 1993, \apj, 418, 244,
  \dodoi{10.1086/173386}

\bibitem[{{Marrone} {et~al.}(2006){Marrone}, {Moran}, {Zhao}, \&
  {Rao}}]{Marrone2006ApJ...640..308M}
{Marrone}, D.~P., {Moran}, J.~M., {Zhao}, J.-H., \& {Rao}, R. 2006, \apj, 640,
  308, \dodoi{10.1086/500106}

\bibitem[{{Mills} {et~al.}(2013){Mills}, {G{\"u}sten}, {Requena-Torres}, \&
  {Morris}}]{2013ApJ...779...47M}
{Mills}, E.~A.~C., {G{\"u}sten}, R., {Requena-Torres}, M.~A., \& {Morris},
  M.~R. 2013, \apj, 779, 47, \dodoi{10.1088/0004-637X/779/1/47}

\bibitem[{{Mossoux} \& {Eckart}(2018)}]{2018MNRAS.474.3787M}
{Mossoux}, E., \& {Eckart}, A. 2018, \mnras, 474, 3787,
  \dodoi{10.1093/mnras/stx3026}

\bibitem[{{Moultaka} {et~al.}(2015{\natexlab{a}}){Moultaka}, {Eckart}, \&
  {Mu{\v{z}}i{\'c}}}]{2015ApJ...806..202M}
{Moultaka}, J., {Eckart}, A., \& {Mu{\v{z}}i{\'c}}, K. 2015{\natexlab{a}},
  \apj, 806, 202, \dodoi{10.1088/0004-637X/806/2/202}

\bibitem[{{Moultaka} {et~al.}(2015{\natexlab{b}}){Moultaka}, {Eckart}, \&
  {Sabha}}]{2015MNRAS.448.3363M}
{Moultaka}, J., {Eckart}, A., \& {Sabha}, N. 2015{\natexlab{b}}, \mnras, 448,
  3363, \dodoi{10.1093/mnras/stv222}

\bibitem[{{Moultaka} {et~al.}(2004){Moultaka}, {Eckart}, {Viehmann}, {Mouawad},
  {Straubmeier}, {Ott}, \& {Sch{\"o}del}}]{Moultaka2004A&A...425..529M}
{Moultaka}, J., {Eckart}, A., {Viehmann}, T., {et~al.} 2004, \aap, 425, 529,
  \dodoi{10.1051/0004-6361:20035807}

\bibitem[{{Mu{\v{z}}i{\'c}} {et~al.}(2010){Mu{\v{z}}i{\'c}}, {Eckart},
  {Sch{\"o}del}, {Buchholz}, {Zamaninasab}, \& {Witzel}}]{2010A&A...521A..13M}
{Mu{\v{z}}i{\'c}}, K., {Eckart}, A., {Sch{\"o}del}, R., {et~al.} 2010, \aap,
  521, A13, \dodoi{10.1051/0004-6361/200913087}

\bibitem[{{Mu{\v{z}}i{\'c}} {et~al.}(2007){Mu{\v{z}}i{\'c}}, {Eckart},
  {Sch{\"o}del}, {Meyer}, \& {Zensus}}]{Muzic2007A&A...469..993M}
{Mu{\v{z}}i{\'c}}, K., {Eckart}, A., {Sch{\"o}del}, R., {Meyer}, L., \&
  {Zensus}, A. 2007, \aap, 469, 993, \dodoi{10.1051/0004-6361:20066265}

\bibitem[{{Najarro} {et~al.}(1997{\natexlab{a}}){Najarro}, {Krabbe}, {Genzel},
  {Lutz}, {Kudritzki}, \& {Hillier}}]{Najarro1997A&A...325..700N}
{Najarro}, F., {Krabbe}, A., {Genzel}, R., {et~al.} 1997{\natexlab{a}}, \aap,
  325, 700

\bibitem[{{Najarro} {et~al.}(1997{\natexlab{b}}){Najarro}, {Krabbe}, {Genzel},
  {Lutz}, {Kudritzki}, \& {Hillier}}]{1997A&A...325..700N}
---. 1997{\natexlab{b}}, \aap, 325, 700

\bibitem[{{Nitschai} {et~al.}(2020){Nitschai}, {Neumayer}, \&
  {Feldmeier-Krause}}]{Nitschai2020ApJ...896...68N}
{Nitschai}, M.~S., {Neumayer}, N., \& {Feldmeier-Krause}, A. 2020, \apj, 896,
  68, \dodoi{10.3847/1538-4357/ab8ea8}

\bibitem[{{Ott}(2012)}]{Ott2012ascl.soft10019O}
{Ott}, T. 2012, {QFitsView: FITS file viewer}.
\newblock \doeprint{1210.019}

\bibitem[{{Ott} {et~al.}(1999){Ott}, {Eckart}, \&
  {Genzel}}]{Ott1999ApJ...523..248O}
{Ott}, T., {Eckart}, A., \& {Genzel}, R. 1999, \apj, 523, 248,
  \dodoi{10.1086/307712}

\bibitem[{{Paumard} {et~al.}(2006){Paumard}, {Genzel}, {Martins}, {Nayakshin},
  {Beloborodov}, {Levin}, {Trippe}, {Eisenhauer}, {Ott}, {Gillessen}, {Abuter},
  {Cuadra}, {Alexander}, \& {Sternberg}}]{Paumard2006ApJ...643.1011P}
{Paumard}, T., {Genzel}, R., {Martins}, F., {et~al.} 2006, \apj, 643, 1011,
  \dodoi{10.1086/503273}

\bibitem[{{Paumard} {et~al.}(2014){Paumard}, {Pfuhl}, {Martins}, {Kervella},
  {Ott}, {Pott}, {Le Bouquin}, {Breitfelder}, {Gillessen}, {Perrin},
  {Burtscher}, {Haubois}, \& {Brandner}}]{Paumard2014A&A...568A..85P}
{Paumard}, T., {Pfuhl}, O., {Martins}, F., {et~al.} 2014, \aap, 568, A85,
  \dodoi{10.1051/0004-6361/201423991}

\bibitem[{{Pei{\ss}ker} {et~al.}(2020{\natexlab{a}}){Pei{\ss}ker}, {Eckart},
  {Sabha}, {Zaja{\v{c}}ek}, \& {Bhat}}]{Peissker2020ApJ...897...28P}
{Pei{\ss}ker}, F., {Eckart}, A., {Sabha}, N.~B., {Zaja{\v{c}}ek}, M., \&
  {Bhat}, H. 2020{\natexlab{a}}, \apj, 897, 28,
  \dodoi{10.3847/1538-4357/ab9826}

\bibitem[{{Pei{\ss}ker} {et~al.}(2020{\natexlab{b}}){Pei{\ss}ker}, {Hosseini},
  {Zaja{\v{c}}ek}, {Eckart}, {Saalfeld}, {Valencia-S.}, {Parsa}, \&
  {Karas}}]{2020A&A...634A..35P}
{Pei{\ss}ker}, F., {Hosseini}, S.~E., {Zaja{\v{c}}ek}, M., {et~al.}
  2020{\natexlab{b}}, \aap, 634, A35, \dodoi{10.1051/0004-6361/201935953}

\bibitem[{{Pei{\ss}ker} {et~al.}(2019){Pei{\ss}ker}, {Zaja{\v{c}}ek}, {Eckart},
  {Sabha}, {Shahzamanian}, \& {Parsa}}]{2019A&A...624A..97P}
{Pei{\ss}ker}, F., {Zaja{\v{c}}ek}, M., {Eckart}, A., {et~al.} 2019, \aap, 624,
  A97, \dodoi{10.1051/0004-6361/201834947}

\bibitem[{{Pei{\ss}ker} {et~al.}(2021){Pei{\ss}ker}, {Ali}, {Zaja{\v{c}}ek},
  {Eckart}, {Hosseini}, {Karas}, {Cl{\'e}net}, {Sabha}, {Labadie}, \&
  {Subroweit}}]{Peissker2021ApJ...909...62P}
{Pei{\ss}ker}, F., {Ali}, B., {Zaja{\v{c}}ek}, M., {et~al.} 2021, \apj, 909,
  62, \dodoi{10.3847/1538-4357/abd9c6}

\bibitem[{{Reid} {et~al.}(2003){Reid}, {Menten}, {Genzel}, {Ott},
  {Sch{\"o}del}, \& {Eckart}}]{Reid2003ApJ...587..208R}
{Reid}, M.~J., {Menten}, K.~M., {Genzel}, R., {et~al.} 2003, \apj, 587, 208,
  \dodoi{10.1086/368074}

\bibitem[{{Rieke} {et~al.}(1978){Rieke}, {Telesco}, \&
  {Harper}}]{1978ApJ...220..556R}
{Rieke}, G.~H., {Telesco}, C.~M., \& {Harper}, D.~A. 1978, \apj, 220, 556,
  \dodoi{10.1086/155936}

\bibitem[{{Roche} {et~al.}(2018){Roche}, {Lopez-Rodriguez}, {Telesco},
  {Sch{\"o}del}, \& {Packham}}]{2018MNRAS.476..235R}
{Roche}, P.~F., {Lopez-Rodriguez}, E., {Telesco}, C.~M., {Sch{\"o}del}, R., \&
  {Packham}, C. 2018, \mnras, 476, 235, \dodoi{10.1093/mnras/sty129}

\bibitem[{{R{\'o}{\.z}a{\'n}ska} {et~al.}(2014){R{\'o}{\.z}a{\'n}ska},
  {Czerny}, {Kunneriath}, {Adhikari}, {Karas}, \&
  {Mo{\'s}cibrodzka}}]{2014MNRAS.445.4385R}
{R{\'o}{\.z}a{\'n}ska}, A., {Czerny}, B., {Kunneriath}, D., {et~al.} 2014,
  \mnras, 445, 4385, \dodoi{10.1093/mnras/stu2066}

\bibitem[{{R{\'o}{\.z}a{\'n}ska} {et~al.}(2017){R{\'o}{\.z}a{\'n}ska},
  {Kunneriath}, {Czerny}, {Adhikari}, \& {Karas}}]{2017MNRAS.464.2090R}
{R{\'o}{\.z}a{\'n}ska}, A., {Kunneriath}, D., {Czerny}, B., {Adhikari}, T.~P.,
  \& {Karas}, V. 2017, \mnras, 464, 2090, \dodoi{10.1093/mnras/stw2460}

\bibitem[{{Sch{\"o}del} {et~al.}(2014){Sch{\"o}del}, {Feldmeier}, {Neumayer},
  {Meyer}, \& {Yelda}}]{2014CQGra..31x4007S}
{Sch{\"o}del}, R., {Feldmeier}, A., {Neumayer}, N., {Meyer}, L., \& {Yelda}, S.
  2014, Classical and Quantum Gravity, 31, 244007,
  \dodoi{10.1088/0264-9381/31/24/244007}

\bibitem[{Sch{\"{o}}del {et~al.}(2009)Sch{\"{o}}del, Merritt, \&
  Eckart}]{Schodel2009}
Sch{\"{o}}del, R., Merritt, D., \& Eckart, A. 2009, Astronomy and Astrophysics,
  502, 91, \dodoi{10.1051/0004-6361/200810922}

\bibitem[{{Sch{\"o}del} {et~al.}(2007){Sch{\"o}del}, {Eckart}, {Alexander},
  {Merritt}, {Genzel}, {Sternberg}, {Meyer}, {Kul}, {Moultaka}, {Ott}, \&
  {Straubmeier}}]{Schodel2007A&A...469..125S}
{Sch{\"o}del}, R., {Eckart}, A., {Alexander}, T., {et~al.} 2007, \aap, 469,
  125, \dodoi{10.1051/0004-6361:20065089}

\bibitem[{{Serabyn} {et~al.}(1991){Serabyn}, {Lacy}, \&
  {Achtermann}}]{Serabyn1991ApJ...378..557S}
{Serabyn}, E., {Lacy}, J.~H., \& {Achtermann}, J.~M. 1991, \apj, 378, 557,
  \dodoi{10.1086/170457}

\bibitem[{{Shukla} {et~al.}(2004){Shukla}, {Yun}, \&
  {Scoville}}]{2004ApJ...616..231S}
{Shukla}, H., {Yun}, M.~S., \& {Scoville}, N.~Z. 2004, \apj, 616, 231,
  \dodoi{10.1086/424868}

\bibitem[{Stolovy {et~al.}(1996)Stolovy, Hayward, \& Herter}]{Stolovy1996}
Stolovy, S.~R., Hayward, T.~L., \& Herter, T. 1996, The Astrophysical Journal,
  470, L45, \dodoi{10.1086/310285}

\bibitem[{Tanner {et~al.}(2002)Tanner, Ghez, Morris, Becklin, Cotera, Ressler,
  Werner, \& Wizinowich}]{Tanner2002}
Tanner, A., Ghez, A.~M., Morris, M., {et~al.} 2002, The Astrophysical Journal,
  575, 860, \dodoi{10.1086/341470}

\bibitem[{{Tsuboi} {et~al.}(2020){Tsuboi}, {Kitamura}, {Tsutsumi}, {Miyawaki},
  {Miyoshi}, \& {Miyazaki}}]{Tsuboi2020PASJ...72...36T}
{Tsuboi}, M., {Kitamura}, Y., {Tsutsumi}, T., {et~al.} 2020, \pasj, 72, 36,
  \dodoi{10.1093/pasj/psaa013}

\bibitem[{{Viehmann} {et~al.}(2005){Viehmann}, {Eckart}, {Sch{\"o}del},
  {Moultaka}, {Straubmeier}, \& {Pott}}]{Viehmann2005A&A...433..117V}
{Viehmann}, T., {Eckart}, A., {Sch{\"o}del}, R., {et~al.} 2005, \aap, 433, 117,
  \dodoi{10.1051/0004-6361:20041748}

\bibitem[{{Viehmann} {et~al.}(2006){Viehmann}, {Eckart}, {Sch{\"o}del}, {Pott},
  \& {Moultaka}}]{Viehmann2006ApJ...642..861V}
{Viehmann}, T., {Eckart}, A., {Sch{\"o}del}, R., {Pott}, J.~U., \& {Moultaka},
  J. 2006, \apj, 642, 861, \dodoi{10.1086/501345}

\bibitem[{{Vollmer} {et~al.}(2004){Vollmer}, {Beckert}, \&
  {Duschl}}]{2004A&A...413..949V}
{Vollmer}, B., {Beckert}, T., \& {Duschl}, W.~J. 2004, \aap, 413, 949,
  \dodoi{10.1051/0004-6361:20034201}

\bibitem[{{Vollmer} \& {Duschl}(2000)}]{Vollmer2000NewA....4..581V}
{Vollmer}, B., \& {Duschl}, W.~J. 2000, \na, 4, 581,
  \dodoi{10.1016/S1384-1076(99)00043-3}

\bibitem[{{Wang} {et~al.}(2013){Wang}, {Nowak}, {Markoff}, {Baganoff},
  {Nayakshin}, {Yuan}, {Cuadra}, {Davis}, {Dexter}, {Fabian}, {Grosso},
  {Haggard}, {Houck}, {Ji}, {Li}, {Neilsen}, {Porquet}, {Ripple}, \&
  {Shcherbakov}}]{2013Sci...341..981W}
{Wang}, Q.~D., {Nowak}, M.~A., {Markoff}, S.~B., {et~al.} 2013, Science, 341,
  981, \dodoi{10.1126/science.1240755}

\bibitem[{{Yusef-Zadeh} \& {Melia}(1992)}]{Yusef-Zadeh1992ApJ...385L..41Y}
{Yusef-Zadeh}, F., \& {Melia}, F. 1992, \apjl, 385, L41, \dodoi{10.1086/186273}

\bibitem[{{Yusef-Zadeh} \& {Morris}(1991)}]{Yusef-Zadeh1991ApJ...371L..59Y}
{Yusef-Zadeh}, F., \& {Morris}, M. 1991, \apjl, 371, L59,
  \dodoi{10.1086/186002}

\bibitem[{{Yusef-Zadeh} {et~al.}(1990){Yusef-Zadeh}, {Morris}, \&
  {Ekers}}]{Yusef-Zadeh1990Natur.348...45Y}
{Yusef-Zadeh}, F., {Morris}, M., \& {Ekers}, R.~D. 1990, \nat, 348, 45,
  \dodoi{10.1038/348045a0}

\bibitem[{{Yusef-Zadeh} {et~al.}(2020){Yusef-Zadeh}, {Royster}, {Wardle},
  {Cotton}, {Kunneriath}, {Heywood}, \& {Michail}}]{2020MNRAS.499.3909Y}
{Yusef-Zadeh}, F., {Royster}, M., {Wardle}, M., {et~al.} 2020, \mnras, 499,
  3909, \dodoi{10.1093/mnras/staa2399}

\bibitem[{{Yusef-Zadeh} {et~al.}(2017){Yusef-Zadeh}, {Wardle}, {Cotton},
  {Sch{\"o}del}, {Royster}, {Roberts}, \&
  {Kunneriath}}]{Yusef-Zadeh2017ApJ...837...93Y}
{Yusef-Zadeh}, F., {Wardle}, M., {Cotton}, W., {et~al.} 2017, \apj, 837, 93,
  \dodoi{10.3847/1538-4357/aa5ea2}

\bibitem[{{Zaja{\v{c}}ek} {et~al.}(2020{\natexlab{a}}){Zaja{\v{c}}ek},
  {Araudo}, {Karas}, {Czerny}, \& {Eckart}}]{2020ApJ...903..140Z}
{Zaja{\v{c}}ek}, M., {Araudo}, A., {Karas}, V., {Czerny}, B., \& {Eckart}, A.
  2020{\natexlab{a}}, \apj, 903, 140, \dodoi{10.3847/1538-4357/abbd94}

\bibitem[{{Zaja{\v{c}}ek} {et~al.}(2020{\natexlab{b}}){Zaja{\v{c}}ek},
  {Araudo}, {Karas}, {Czerny}, {Eckart}, {Sukov{\'a}}, {{\v{S}}tolc}, \&
  {Witzany}}]{2020arXiv201112868Z}
{Zaja{\v{c}}ek}, M., {Araudo}, A., {Karas}, V., {et~al.} 2020{\natexlab{b}},
  arXiv e-prints, arXiv:2011.12868.
\newblock \doarXiv{2011.12868}

\bibitem[{{Zhao} {et~al.}(2010){Zhao}, {Blundell}, {Moran}, {Downes},
  {Schuster}, \& {Marrone}}]{2010ApJ...723.1097Z}
{Zhao}, J.-H., {Blundell}, R., {Moran}, J.~M., {et~al.} 2010, \apj, 723, 1097,
  \dodoi{10.1088/0004-637X/723/2/1097}

\bibitem[{{Zhao} {et~al.}(2009){Zhao}, {Morris}, {Goss}, \&
  {An}}]{2009ApJ...699..186Z}
{Zhao}, J.-H., {Morris}, M.~R., {Goss}, W.~M., \& {An}, T. 2009, \apj, 699,
  186, \dodoi{10.1088/0004-637X/699/1/186}

\end{thebibliography}
\bibliographystyle{aasjournal}

\appendix

\section{{Tangential velocity} comparisons between MIR and K-band velocities.}
\label{sec:ProperMotion_Comp}

The comparison between the K-band {tangential velocity} (\cite{Schodel2009} and \cite{Genzel2000MNRAS.317..348G}) and our mid-infrared velocities for the
point sources (see Table \ref{tab:Comparision_pm}) looks favourable. {The {mean absolute difference} of around $\pm$100~km/s is most likely affected by the larger point spread function in the MIR, the limited baseline in time, and the fact the even the
point sources may show some IR excess/extension or they are located on background emission that is spatially structured/variable on scales of the point spread function. }

\begin{table*}[h]
\renewcommand\thetable{5}
\caption{Comparison of stellar {tangential velocities} in PAH1 and NeII\_2 bands with those in K-band. All velocities are in $kms^{-1}$. S09 is the data from \cite{Schodel2009}. G00 is the {data} from \cite{Genzel2000MNRAS.317..348G}. Sources used as calibrators are marked with *. We choose a conservative 0.25 pixel uncertainties for PAH1 and NeII\_2 proper motions which correspond to about 45$ km/s$. }
\label{tab:Comparision_pm}
\centering
\begin{tabular}{c	 r r		 r r		 r r r r      r r r  r      r r r r}
\hline
\hline
    & \multicolumn{1}{|l}{}  &PAH1 &\multicolumn{1}{|l}{}&Ne II&\multicolumn{1}{|l}{}&K&&&\multicolumn{1}{|l}{}&S09&&&\multicolumn{1}{|l}{}&G00&&\\
Source & \multicolumn{1}{|r}{$v_{\alpha}$} &  $v_{\delta}$  &
 \multicolumn{1}{|r}{$v_{\alpha}$} &  $v_{\delta}$ &  \multicolumn{1}{|r}{$v_{\alpha}$}  &  $\Delta v_{\alpha}$ & $v_{\delta}$ & $ \Delta v_{\delta}$&  \multicolumn{1}{|r}{$v_{\alpha}$} &  $ \Delta v_{\alpha}$ & $v_{\delta}$ & $\Delta v_{\delta}$& \multicolumn{1}{|r}{$v_{\alpha}$} &  $ \Delta v_{\alpha}$ &  $v_{\delta}$ & $\Delta v_{\delta}$\\ 

\hline	
\multicolumn{1}{l|}{	IRS5	}&	-303	&	167	&	\multicolumn{1}{|r}{	-71	}&	-145	&	\multicolumn{1}{|r}{	-267	}&	6	 &	84	& 	12	&	\multicolumn{1}{|r}{		}&	    	&	    	&	    	&	\multicolumn{1}{|r}{		}&	    	&		&		\\
\multicolumn{1}{l|}{	IRS10W	}&	-118	&	331	&	\multicolumn{1}{|r}{	120	}&	148	&	\multicolumn{1}{|r}{	-29	}&	6	&	115	&	10	&	\multicolumn{1}{|r}{	    	}&	    	&	    	&	    	&	\multicolumn{1}{|r}{		}&	    	&		&		\\
\multicolumn{1}{l|}{	IRS10EE*	}&	-157	&	54	&	\multicolumn{1}{|r}{	-186	}&	142	&	\multicolumn{1}{|r}{	-44	}&	6	&	-16	&	10	&	\multicolumn{1}{|r}{	-15	}&	7	&	-60	&	7	&	\multicolumn{1}{|r}{		}&		&		&		\\
\multicolumn{1}{l|}{	IRS1W	}&	-113	&	307	&	\multicolumn{1}{|r}{	53	}&	275	&	\multicolumn{1}{|r}{	-9	}&	6	&	131	&	10	&	\multicolumn{1}{|r}{	    	}&	    	&	    	&	    	&	\multicolumn{1}{|r}{		}&		&		&		\\
\multicolumn{1}{l|}{	IRS16NE	}&	-83	&	-224	&	\multicolumn{1}{|r}{	    	}&	    	&	\multicolumn{1}{|r}{	104	}&	6	&	-281	&	10	&	\multicolumn{1}{|r}{	    	}&	    	&	    	&	    	&	\multicolumn{1}{|r}{	199	}&	65	&	-279	&	21	\\
\multicolumn{1}{l|}{	IRS16C	}&	-513	&	358	&	\multicolumn{1}{|r}{	    	}&	    	&	\multicolumn{1}{|r}{	-211	}&	6	&	108	&	10	&	\multicolumn{1}{|r}{	    	}&	    	&	    	&	    	&	\multicolumn{1}{|r}{	-330	}&	39	&	353	&	34	\\
\multicolumn{1}{l|}{	IRS21	}&	-138	&	8	&	\multicolumn{1}{|r}{	-65	}&	3	&	\multicolumn{1}{|r}{	-1	}&	6	&	-43	&	10	&	\multicolumn{1}{|r}{	    	}&	    	&	    	&	    	&	\multicolumn{1}{|r}{	-159	}&	65	&	64	&	38	\\
\multicolumn{1}{l|}{	IRS9*	}&	97	&	-22	&	\multicolumn{1}{|r}{	    	}&	    	&	\multicolumn{1}{|r}{	176	}&	6	&	90	&	10	&	\multicolumn{1}{|r}{	127	}&	10	&	116	&	10	&	\multicolumn{1}{|r}{		}&		&		&		\\
\multicolumn{1}{l|}{	IRS2S	}&	180	&	-124	&	\multicolumn{1}{|r}{	82	}&	-230	&	\multicolumn{1}{|r}{	352	}&	8	&	-368	&	14	&	\multicolumn{1}{|r}{	    	}&	    	&	    	&	    	&	\multicolumn{1}{|r}{		}&		&		&		\\
\multicolumn{1}{l|}{	IRS2L	}&	173	&	-2	&	\multicolumn{1}{|r}{	293	}&	191	&	\multicolumn{1}{|r}{	174	}&	8	&	-273	&	10	&	\multicolumn{1}{|r}{	    	}&	    	&	    	&	    	&	\multicolumn{1}{|r}{		}&		&		&		\\
\multicolumn{1}{l|}{	IRS29	}&	256	&	-186	&	\multicolumn{1}{|r}{	585	}&	-205	&	\multicolumn{1}{|r}{	165	}&	6	&	-261	&	10	&	\multicolumn{1}{|r}{	    	}&	    	&	    	&	    	&	\multicolumn{1}{|r}{		}&		&		&		\\
\multicolumn{1}{l|}{	IRS29NE	}&	-435	&	7	&	\multicolumn{1}{|r}{	    	}&	    	&	\multicolumn{1}{|r}{	-215	}&	6	&	-151	&	10	&	\multicolumn{1}{|r}{	    	}&	    	&	    	&	    	&	\multicolumn{1}{|r}{		}&		&		&		\\
\multicolumn{1}{l|}{	IRS34	}&	119	&	166	&	\multicolumn{1}{|r}{	    	}&	    	&	\multicolumn{1}{|r}{	-24	}&	8	&	-335	&	10	&	\multicolumn{1}{|r}{	    	}&	    	&	    	&	    	&	\multicolumn{1}{|r}{		}&		&		&		\\
\multicolumn{1}{l|}{	IRS6E	}&	294	&	234	&	\multicolumn{1}{|r}{	398	}&	-206	&	\multicolumn{1}{|r}{	141	}&	6	&	-39	&	10	&	\multicolumn{1}{|r}{	    	}&	    	&	    	&	    	&	\multicolumn{1}{|r}{		}&		&		&		\\
\multicolumn{1}{l|}{	IRS3	}&	282	&	-2	&	\multicolumn{1}{|r}{	412	}&	-13	&	\multicolumn{1}{|r}{	84	}&	8	&	-137	&	10	&	\multicolumn{1}{|r}{	    	}&	    	&	    	&	    	&	\multicolumn{1}{|r}{	170	}&	40	&	115	&	45	\\
\multicolumn{1}{l|}{	IRS7*	}&	27	&	-125	&	\multicolumn{1}{|r}{	232	}&	-250	&	\multicolumn{1}{|r}{	-23	}&	8	&	-193	&	12	&	\multicolumn{1}{|r}{	-2	}&	9	&	-176	&	9	&	\multicolumn{1}{|r}{	100	}&	67	&	-118	&	35	\\
\multicolumn{1}{l|}{	IRS12N*	}&	80	&	-133	&	\multicolumn{1}{|r}{	-5	}&	-106	&	\multicolumn{1}{|r}{	    	}&	    	&	    	&	    	&	\multicolumn{1}{|r}{	-62	}&	8	&	-107	&	8	&	\multicolumn{1}{|r}{		}&		&		&		\\
\multicolumn{1}{l|}{	IRS15NE*	}&	   -67 	&	    -133	&	\multicolumn{1}{|r}{	-6	}&	-308	&	\multicolumn{1}{|r}{	    	}&	    	&	    	&	    	&	\multicolumn{1}{|r}{	-58	}&	6	&	-223	&	6	&	\multicolumn{1}{|r}{		}&		&		&		\\
\multicolumn{1}{l|}{	IRS17*}&	-351&	198 &	\multicolumn{1}{|r}{	-238	}&	-83	&	\multicolumn{1}{|r}{	    	}&	    	&	    	&	    	&	\multicolumn{1}{|r}{	-65	}&	5	&	-40	&	5	&	\multicolumn{1}{|r}{		}&		&		&		\\

\hline
\end{tabular}
\end{table*}

\begin{figure}[h]
\begin{center}
\includegraphics[width=0.6\textwidth, trim = 0 45 0 0, clip]{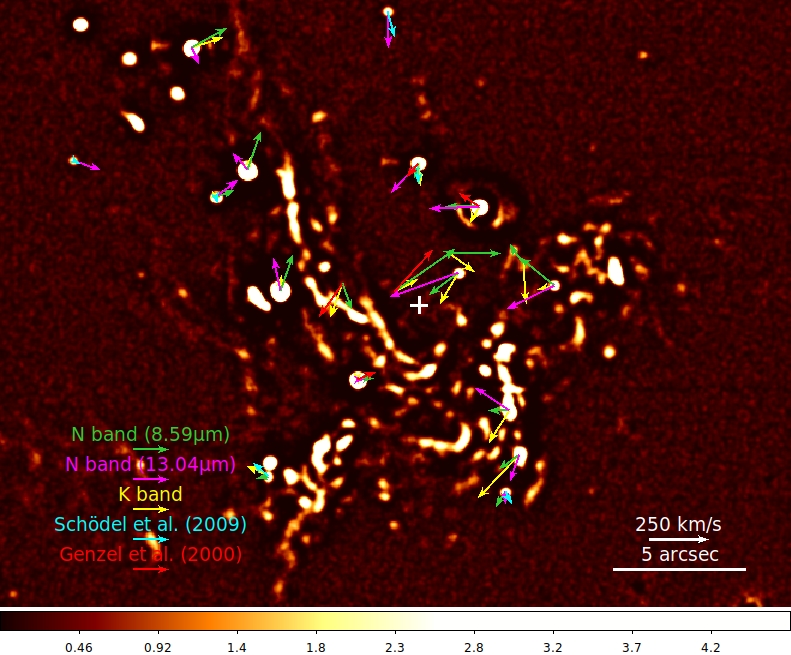}
\caption{Comparison velocities from K-band and \cite{Schodel2009} and \cite{Genzel2000MNRAS.317..348G}}.
\label{fig:Comparision_pm}
\end{center}
\end{figure}

\begin{figure}[!htb]

\minipage{0.33\textwidth}
  \includegraphics[width=\linewidth, trim = 10 0 30 0]{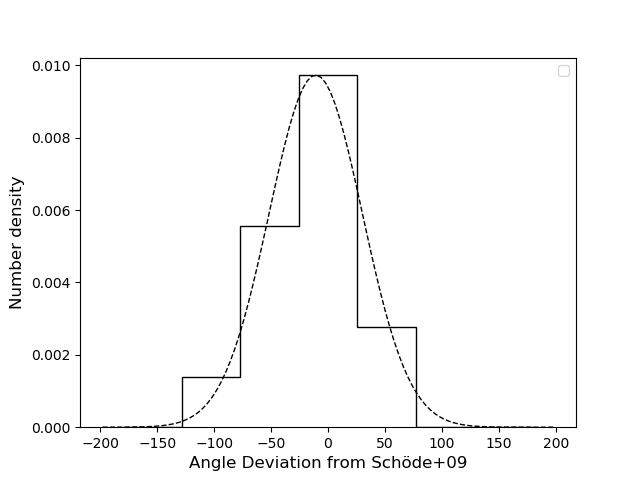}
\endminipage\hfill
\minipage{0.33\textwidth}
  \includegraphics[width=\linewidth, trim = 10 0 30 0]{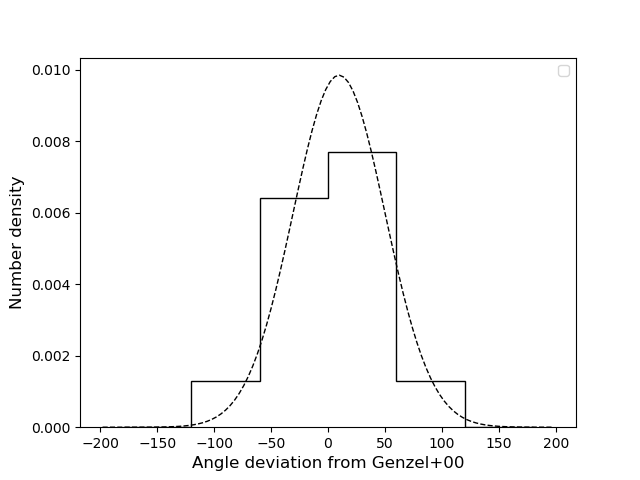}
\endminipage\hfill
\minipage{0.33\textwidth}%
  \includegraphics[width=\linewidth, trim = 10 0 30 0]{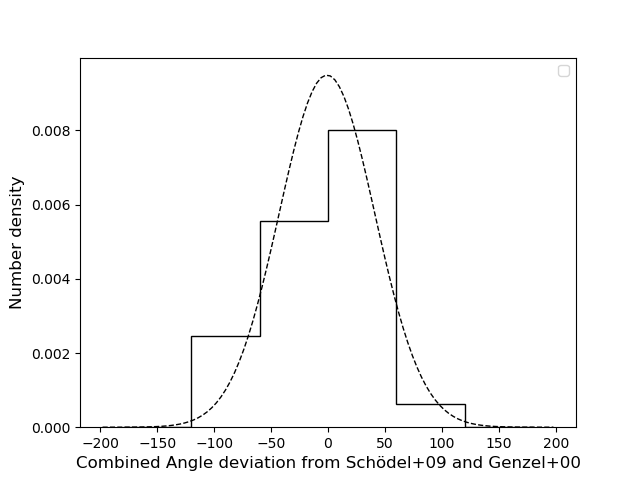}
\endminipage
\caption{Angle deviations from \cite{Schodel2009} and \cite{Genzel2000MNRAS.317..348G} values in all the wavelengths. Mean absolute deviation in Velocity(RA) is about 102km/s, in Velocity(dec) is about 76 km/s and standard deviation of the angle difference is about 40 degrees. }
\label{fig:Deviation}
\end{figure}
\newpage
\section{$N$-band Flux Density Comparison}
\label{sec:Flux_Comparison}

\begin{table*}[h]
\renewcommand\thetable{6}
\caption{Comparison of measured flux densities (reddened) with $N$-band fluxes of \cite{Viehmann2006ApJ...642..861V}.
Uncertainties in V+06 flux densities are $\pm$ 30$\%$. Sources used as flux calibrators are marked with *.}
\label{tab:Flux_Comparions}
\centering
\begin{tabular}{l r r r r r r r r}
\hline
\hline
\multicolumn{1}{r|}{}&&PAH1&&&\multicolumn{1}{|r}{}& NeII\_2 &&	\\
\multicolumn{1}{c|}{Source}& \multicolumn{1}{c}{F (Jy)}& \multicolumn{1}{c}{F (Jy)}& \multicolumn{1}{c}{F (Jy)}&	\multicolumn{1}{c}{$\pm$F (Jy)}& \multicolumn{1}{|c}{F (Jy)}&  \multicolumn{1}{c}{F (Jy)}& \multicolumn{1}{c}{F (Jy)}& \multicolumn{1}{c}{$\pm$F (Jy)}\\
\multicolumn{1}{r|}{}& \multicolumn{1}{c}{(V+06)}	& \multicolumn{1}{c}{(Circular)} & \multicolumn{1}{c}{(FWHM)} &	 &\multicolumn{1}{|c}{(V+06)	}& \multicolumn{1}{c}{(Circular)}	& \multicolumn{1}{c}{(FWHM)}  	&	\\
\hline	

\multicolumn{1}{l|}{	IRS 5NE*	}&	0.51	&	0.59	&	0.57	&	0.09	&\multicolumn{1}{|r}{	0.56	}&	0.50	&	1.00	&	0.09	\\
\multicolumn{1}{l|}{	IRS 5E	}&	0.59	&	0.38	&	0.49	&	0.08	&\multicolumn{1}{|r}{	1.54	}&	0.57	&	1.99	&	0.17	\\
\multicolumn{1}{l|}{	IRS 5S	}&	0.63	&	0.25	&	0.61	&	0.09	&\multicolumn{1}{|r}{	0.38	}&	0.29	&	2.35	&	0.20	\\
\multicolumn{1}{l|}{	IRS 5	}&	5.11	&	5.44	&	3.92	&	0.60	&\multicolumn{1}{|r}{	5.85	}&	4.78	&	4.54	&	0.38	\\
\multicolumn{1}{l|}{	IRS 10W*	}&	11.06	&	10.18	&	8.33	&	1.27	&\multicolumn{1}{|r}{	11.85	}&	10.94	&	11.61	&	0.98	\\
\multicolumn{1}{l|}{	IRS 7*	}&	1.47	&	1.26	&	1.15	&	0.18	&\multicolumn{1}{|r}{	1.75	}&	1.95	&	2.04	&	0.17	\\
\multicolumn{1}{l|}{	IRS 3	}&	13.42	&	16.35	&	8.65	&	1.31	&\multicolumn{1}{|r}{	12.85	}&	11.17	&	11.75	&	0.99	\\
\multicolumn{1}{l|}{	IRS 1W*	}&	20.42	&	22.96	&	16.14	&	2.45	&\multicolumn{1}{|r}{	22.93	}&	25.40	&	21.97	&	1.85	\\
\multicolumn{1}{l|}{	IRS 21	}&	4.56	&	4.95	&	4.79	&	0.73	&\multicolumn{1}{|r}{	5.44	}&	6.27	&	6.92	&	0.58	\\
\multicolumn{1}{l|}{	IRS 29	}&	0.15	&	0.07	&	0.13	&	0.02	&\multicolumn{1}{|r}{		}&		&		&		\\
\multicolumn{1}{l|}{	IRS 2L	}&	5.26	&	4.25	&	3.28	&	0.50	&\multicolumn{1}{|r}{	7.49	}&	6.45	&	5.52	&	0.47	\\
\multicolumn{1}{l|}{	IRS 9	}&	0.21	&	0.58	&	0.39	&	0.06	&\multicolumn{1}{|r}{		}&		&		&		\\

\hline
\end{tabular}
\end{table*}



\end{document}